\documentclass[12pt]{article} 
\usepackage{amsmath,amssymb} 

\newtheorem{theorem}{Theorem}[section]
\newcommand{\proof}{{\noindent \it Proof:\ }}
\newcommand{\qed}{\hfill $\Box$\par\hfill\\}

\newtheorem{proposition}{Proposition}[section]
\newtheorem{lemma}{Lemma}[section]
\newtheorem{corollary}{Corollary}[section]
\newtheorem{remark}{Remark}

\numberwithin{equation}{section}

\begin{document}

\title{Infrared Catastrophe for Nelson's Model
\thanks{to appear in \textit{Publ. RIMS}}}

\author{Masao Hirokawa\thanks{This work is supported by JSPS, 
Grant-in-Aid for Scientific Research (C) 16540155.}
}

\date{}
\maketitle

\abstract{
We mathematically study the infrared catastrophe 
for the Hamiltonian of Nelson's model when it has 
the external potential in a general class.  
For the model, we prove the pull-through formula 
on ground states in operator theory first. 
Based on this formula, we show both non-existence of 
any ground state and divergence of the total number 
of soft bosons.  
}

\section{Introduction} 
\label{section:Intro}

The purpose of the present paper is to investigate 
mathematically the infrared (IR) catastrophe 
for Nelson's Hamiltonian \cite{nelson}, 
in particular non-existence of ground state 
and the divergence of the total number 
of soft bosons ({\it soft-boson divergence}). 
The exact definition of ground state 
will be stated in \S \ref{section:main}. 
The definition of soft boson will be explained later.   
IR catastrophe is the trouble of IR divergence 
caused by {\it massless} particles forming 
a quantized field.  
Nelson's Hamiltonian is the Hamiltonian 
of the so-called Nelson's model 
describing a system of a quantum particle, which moves 
in the $3$-dimensional Euclidean space ${\mathbb R}^{3}$ 
under the influence of an external potential, 
and which interacts with a massless scalar Bose field.  
The massless scalar Bose field is the quantized scalar field 
made of {\it massless bosons}. 
The boson is the (quantum) particle following the 
Bose-Einstein statistics. 
In the present paper the soft boson means the 
boson in a ground state.

Recently, the spectral properties of Nelson's Hamiltonian 
has been studied rather intensively 
(e.g., \cite{arai,betz,dg,hhs2,hhs,lorinczi}). 
In particular, Betz {\it et al}. showed in \cite{betz} 
that when the external potential is in the Kato class 
the total number of soft bosons 
for Nelson's Hamiltonian 
diverges under the {\it infrared singularity} 
(IRS) condition. 
We will concretely define this condition 
in \S \ref{section:main}. 
Around the same time L\H{o}rinczi {\it et al}. 
showed in \cite{lorinczi} that 
when the external potential is strongly confining
there is no ground state of Nelson's Hamiltonian 
in spatial dimension $3$. 
The results in both \cite{betz} and \cite{lorinczi} 
are proved by means of functional integrals. 

In \cite{dg} Deresi\'{n}ski and G\'{e}rard treated 
the problem of non-existence of ground state 
by $L^{2}$-theoretical method and 
proved the non-existence of any ground state 
for Nelson's Hamiltonian under the assumption 
that the external potential is strongly confining. 
They employed an amazingly simple method based on the 
$L^{2}$-theoretical pull-through formula. 
However, the results shown in \cite{dg} 
do not seem to include the case of decaying potentials 
such as the Coulomb potential. 
For another model, the so-called Pauli-Fierz model 
\cite{pf}, 
it was clarified in \cite{bfs2, gll} 
that there exists a ground state 
even under IRS condition, 
when Pauli-Fierz's Hamiltonian 
has the Coulomb-type potential. 

In the present paper we consider Nelson's Hamiltonian 
with a general class of potentials including 
both strongly confining potentials and 
Coulomb-type potentials and 
prove in a unified way the non-existence of any 
ground state and the soft-boson divergence.  
Following the methods in \cite{dg,lorinczi} 
to prove the non-existence of any ground state,  
we are required to invent some suitable technique 
in order to include Coulomb-type potentials. 
Thus, the present paper looks at the problem 
from a different angle. 
Following the physical observation stated below, 
we adopt an operator-theoretical method in which 
we combine the technique of spatial localization 
presented by Griesemer, Lieb, and Loss \cite{gll} 
and an approach based on the proof of the absence 
of ground state by Arai, Hiroshima, and the 
author \cite{ahh}. 
We believe that this approach is new. 

In this paper 
the operator-theoretical pull-through formula 
announced in \cite{hirokawa-nelson'} 
plays a crucial role. 
So, we give a complete version of its proof. 
To the best of author's knowledge, 
the approach presented in this paper is the first 
to establish the pull-through formula in an 
operator-theoretical framework. 
Such an operator-theoretical formula makes it possible 
to analyze infrared catastrophe in mathematical detail 
\cite{ahh2,hirokawa-IR, hiroshima}. 
In physics it is generally expected 
that the non-existence of ground state results from 
the soft-boson divergence. 
From a mathematical point of view, however, 
we establish in the present paper that 
the pull-through formula implies both 
the non-existence of ground state 
(Theorems \ref{theorem:2} and  \ref{crl:2}) 
and the soft-boson divergence 
(Theorems \ref{theorem:1} and \ref{crl:1}), 
independently to each other.
 
In a mathematical treatment, this IR problem was 
first studied for a fermion-boson model related to 
Nelson's by Fr\"{o}hlich \cite{fr}. 
It is worthy of note that Pizzo developed 
Fr\"{o}hlich's work in \cite{pi}.    
We tackled IR problem of proving the non-existence 
of ground state for the so-called generalized spin-boson 
(GSB) model from an operator-theoretical 
point of view in \cite{ahh},  
while we studied a mathematical mechanism 
of existence of ground states for it in \cite{ah1}. 
However, because GSB model is very general, 
the information on IR problem for it was 
so limited that we could not entirely achieve our goal. 
In the present paper, we completely achieve it 
for Nelson's Hamiltonian with the external potential 
in the general class. 

For our goal, we present the following physical image 
of the relation between the soft-boson divergence 
and the non-existence of ground state: 
To begin with, the quantum particle coupled with 
the field formed by bosons is generally dressed in 
the cloud of bosons, which makes the so-called 
{\it quasi-particle}. 
In particular, the total number of soft bosons 
for Nelson's model diverges under IRS condition. 
So, if a ground state exists under IRS condition, 
then the quantum particle has to dress itself in 
the cloud of infinitely many soft bosons. 
Thus, we can hardly expect that the cloud is spatially 
localized into a finite area. 
Namely, because the soft boson is the boson in a ground state, 
the uncertainty of the particle's position 
in the ground state must be infinite under IRS condition. 
On the other hand, once a ground state exists, 
we can generally expect to obtain the finite uncertainty of 
the position in the ground state in order to observe 
the particle's position. 
Therefore, the existence of a ground state of 
Nelson's model under IRS condition must imply 
a contradiction in quantum theory. 
We seek to express this image in a mathematical way.  

The present paper is organized as follows. 
In \S \ref{section:main} we state main results. 
On the external potential we impose two kind of assumptions, 
assumption (A) and assumption (C). 
The assumption (A) is of rather general nature. 
Assuming (A), we assert that ground states are 
absent from the domain of the square of position 
operator (Theorem \ref{theorem:2}). 
Assumption (C) is more concrete and more restrictive 
than (A). Assuming (C), we establish the non-existence 
of any ground state (Theorem \ref{theorem:1}). 
Theorems \ref{crl:2} and \ref{crl:1} 
are concerned with estimates of 
number of soft bosons. 
In \S \ref{section:RPTF} the operator-theoretical 
pull-through formula 
is proved and a useful identity is derived from it. 
In \S \ref{section:ABG} we prove Theorem \ref{theorem:2} 
and in \S \ref{section:EESPB} Theorem \ref{theorem:1}. 
In \S \ref{section:FNPGS} the finite uncertainty is argued, 
and combining this with the absence theorem and the estimate 
proved in \S \ref{section:ABG}, 
we establish our final results, 
Theorems \ref{crl:2} and \ref{crl:1}.

\section{Main results}
\label{section:main}

The position of the quantum particle with mass 
$m = 1$ is denoted by $x$, 
the momentum by $p := -i\nabla_{x}$. 
Here we employ the natural units. 
Namely, we set $\hbar=1, c=1$ throughout. 
As the Hamiltonian for the quantum particle, 
we consider the Schr\"{o}dinger operator 
acting in $L^{2}({\mathbb R}^{3})$, 
$${H_{\mathrm {at}}} :=  
\frac{1}{2}p^2  + V,$$ 
with an external potential $V$.  

We consider two types of assumption for $H_{\mathrm {at}}$ 
as the notice was given in \S \ref{section:Intro}, 
i.e., general assumption (A) and concrete assumption (C). 
We prove under (A) that any ground state is not in 
the subspace characterized by a kind of spatial localization 
(Theorem \ref{theorem:2}). 
Under (C) we completely prove the non-existence of 
any ground state (Theorem \ref{crl:2}). 

\begin{description}
\item[(A)] $H_{\mathrm{at}}$ is a self-adjoint operator 
bounded from below such that 
$D(H_{\mathrm {at}}) \subset D(p^{2})$. 
Moreover, $H_{\mathrm{at}}$ has a 
ground state $\psi_{\mathrm {at}}$.  
\end{description}
Here $D(T)$ denotes the domain of an operator $T$.  
We denote the ground state energy by $E_{\mathrm {at}} 
:= \inf\sigma({H_{\mathrm {at}}})$, 
where $\sigma (T)$ denotes the spectrum of a closed operator $T$.

For completion of the non-existence theorem, 
we investigate the following two classes 
of external potentials. 
The two classes include 
the strongly confining potential, long and 
short range ones.

\begin{description}
\item[(C1)]\cite{arai}: 
\item[(C1-1)] $H_{\mathrm {at}}$ is self-adjoint 
on $D(H_{\mathrm {at}}) 
= D(p^{2})\cap D(V)$ and bounded from below, 
\item[(C1-2)] there exist positive constants 
$c_{1}$ and $c_{2}$ such that 
$|x|^{2} 
\le c_{1}V(x) + c_{2}$ 
for almost every (a.e.) $x \in  {\mathbb R}^{3}$, and 
${\displaystyle \int_{|x| \le R}|V(x)|^{2}d^{3}x < \infty}$  
for all $R > 0$.  
\end{description}

\begin{description}
\item[(C2)]  \cite{spohn}:  
\item[(C2-1)] $V \in  L^{2}({\mathbb R}^{3}) + 
L^{\infty}({\mathbb R}^{3})$, and 
$\lim_{|x|\to\infty}|V(x)| = 0$. 
\end{description}
In this case, 
by Kato's theorem \cite[Theorem X15]{rs2} and 
the well-known fact \cite[\S XIII.4, Example 6]{rs4}, 
we have the following: 

\begin{proposition}
\label{proposition:schroedinger-sa} 
Assume (C2-1). Then, 
\begin{description}
\item[(i)]
${H_{\mathrm {at}}}$ is self-adjoint on $D(p^{2})$. 
\item[(ii)] $V$ is infinitesimally $p^{2}$-bounded. 
\item[(iii)] $\sigma_{\mathrm {ess}}(H_{\mathrm {at}}) = 
\left[\left. 0\, ,\, \infty \right)\right.$, where 
$\sigma_{\mathrm {ess}}(H_{\mathrm {at}})$ is the 
essential spectrum of $H_{\mathrm {at}}$. 
\end{description}
\end{proposition}
We assume the following in addition to (C2-1): 
\begin{description}
\item[(C2-2)] $H_{\mathrm {at}}$ has a ground state 
$\psi_{\mathrm {at}}$ satisfying 
${\psi_{\mathrm {at}}} (x) > 0$ for  
a.e. $x \in {\mathbb R}^{3}$ and  
$E_{\mathrm {at}} < 0$.  
\end{description}

Both in (C1) and (C2), condition (A) holds and 
we have a ground state $\psi_{\mathrm {at}}$ 
of $H_{\mathrm {at}}$. 
We say that {\it $V$ is in (C1)} (resp. {\it (C2)}) 
if (C1-1) and (C1-2) (resp. (C2-1) and (C2-2)) 
hold.

Our quantum particle is coupled 
with a massless scalar Bose field. 
We first prepare some notations for the quantized field. 
For the state space of scalar bosons, we take 
the Hilbert space given by the symmetric Fock space 
${\mathcal F} := \bigoplus_{n=0}^{\infty}
\left[ \otimes_{\mathrm {s}}^{n}L^{2}({\mathbb R}^{3})\right]$ 
over $L^{2}({\mathbb R}^{3})$, where 
$\otimes_{\mathrm {s}}^{n}L^{2}({\mathbb R}^{3})$ 
denotes the $n$-fold symmetric tensor product 
of $L^{2}({\mathbb R}^{3})$, the space of all 
square-integrable functions, and 
$\otimes_{\mathrm {s}}^{0}L^{2}({\mathbb R}^{3}) 
:= {\mathbb C}$. 
The finite particle space ${\mathcal F}_{0}$ is 
defined by  ${\mathcal F}_{0}$ 
$:=$ $\left\{\right.$ $\Psi$ $=$ 
$\Psi^{(0)}\oplus\cdots\oplus
\Psi^{(n)}\oplus\cdots$ $\in$  ${\mathcal F}$ $|$ 
$\Psi^{(n)} = 0$ for $n \ge \exists n_{0}$ 
$\left.\right\}$.  
For every $f \in L^{2}({\mathbb R}^{3})$ and 
$\Psi = \Psi^{(0)}\oplus\Psi^{(1)}\oplus
\cdots\oplus\Psi^{(n)}\oplus\cdots \in  
{\mathcal F}_{0}$, 
the smeared annihilation operator $a(f)$ 
of bosons is defined by 
\begin{eqnarray} 
\qquad \left( a(f)\Psi\right)^{(n)}(k_{1}, \cdots, k_{n}) 
:= \sqrt{n+1}\int_{{\mathbb R}^{3}} 
f(k)^{*}\Psi^{(n+1)}(k, k_{1}, \cdots, k_{n})d^{3}k
\label{eq:smeard-annihilation}
\end{eqnarray}
as $\otimes_{\mathrm s}^{n+1}L^{2}({\mathbb R}^{3}) \ni 
\Psi^{(n+1)} \to \left( a(f)\Psi\right)^{(n)} 
\in  \otimes_{\mathrm s}^{n}L^{2}({\mathbb R}^{3})$ 
for $n = 0, 1, 2, \cdots$, 
where $f(k)^{*}$ is the complex conjugate of $f 
\in L^{2}({\mathbb R}^{3})$.   
Then, $a(f)$ is closable for every 
$f \in  L^{2}({\mathbb R}^{3})$. 
We denote its closure by the same symbol. 
We define the smeared creation operator $a^{\dagger}(f)$ 
by the adjoint operator of $a(f)$, i.e., 
$a^{\dagger}(f) = a(f)^{*}$, for every 
$f \in L^{2}({\mathbb R}^{3})$.

The smeared annihilation and creation operators 
satisfy the standard canonical commutation relations 
(CCR): 
$$[a(f),a^{\dagger}(g)]= (f,g)_{L^{2}} 
\equiv \int_{{\mathbb R}^{3}}f(k)^{*}g(k)d^{3}k,$$
$$[a(f),a(g)]=0,\ \ \ [a^{\dagger}(f),a^{\dagger}(g)]=0,\ \ \ 
\forall f, g\in L^{2}({\mathbb R}^{3}),$$ 
on ${\mathcal F}_{0}$.

In this paper, we consider the following 
dispersion relation $\omega(k)$, 
\begin{eqnarray} 
\omega(k) = |k|.
\label{eq:dispersion}
\end{eqnarray}
Then the free field energy operator 
$H_{\mathrm f}$ is the second quantization 
of $\omega$, i.e., 
\begin{eqnarray*}
H_{\mathrm f} := d\Gamma(\omega). 
\end{eqnarray*}
Here, for a self-adjoint operator $h$ acting 
in $L^{2}({\mathbb R}^{3})$, its second quantization 
is defined by 
\begin{eqnarray*}
d\Gamma (h) := \bigoplus_{n=0}^{\infty}h^{(n)},
\end{eqnarray*}  
where $h^{(n)}$ is the closure of $\sum_{j=1}^{n} 
I\otimes \cdots\otimes 
{\displaystyle 
\mathop{\mathop{h}^{\smile}}^{j\mbox{\rm {\footnotesize -th}}}}
\otimes\cdots\otimes I 
\equiv h\otimes I\otimes\cdots\otimes I 
+ I\otimes h\otimes I\otimes\cdots\otimes I 
+ \cdots + I\otimes\cdots\otimes I\otimes h$, 
i.e., 
$$h^{(n)} := \overline{\sum_{j=1}^{n} 
I\otimes \cdots\otimes 
\mathop{\mathop{h}_{\frown}}_{j\mbox{\rm {\footnotesize -th}}}
\otimes\cdots\otimes I}$$
acting in $\otimes_{\mathrm s}^{n}L^{2}({\mathbb R}^{3})$,  
where $I$ denotes the identity operator 
on $L^{2}({\mathbb R}^{3})$, and $h^{(0)} = 0$. 
We note that $d\Gamma(h)$ is a self-adjoint operator 
acting in ${\mathcal F}$. 
Thus, for $H_{\mathrm {f}}$ 
we employed the multiplication operator $\omega$ 
as $h$ in (\ref{eq:dispersion}). 
We define the subspace 
${\mathcal F}(\omega)$ by the linear hull 
of $\left\{\right.$ $\Omega_{0}, a^{\dagger}(f_{1})\cdots 
a^{\dagger}(f_{\nu})\Omega_{0}\, |\,$ 
$\nu \in  {\mathbb N},$ $f_{j} \in  D(\omega),$ $j = 1, \cdots, 
\nu$ $\left.\right\}$, 
where $\Omega_{0}$ is the Fock vacuum, i.e., 
\begin{eqnarray*}
\Omega_{0} = 1\oplus 0 \oplus 0 \oplus\cdots \in  {\mathcal F}.
\end{eqnarray*}
Then, the action of $H_{\mathrm f}$ is given by 
$$\otimes_{\mathrm s}^{n}L^{2}({\mathbb R}^{3}) \ni 
\left( H_{\mathrm f}\Psi\right)^{n}(k_{1}, \cdots, k_{n}) 
= \sum_{j=1}^{n}|k_{j}| 
\Psi^{(n)}(k_{1}, \cdots, k_{n}),  
\quad \forall n \in  {\mathbb N},$$ 
and $\left( H_{\mathrm f}\Psi\right)^{(0)} = 0$ 
for $\Psi = \Psi^{(0)}\oplus\Psi^{(1)}\oplus \cdots$ 
$\in$  ${\mathcal F}(\omega)$. 
$H_{\mathrm f}$ is symbolically written as 
$$ H_{\mathrm f}  
= \int_{{\mathbb R}^{3}} 
|k| a^{\dagger}(k) a(k)d^{3}k, 
$$ 
using symbolical representation of 
the annihilation operator by the kernel $a(k)$,   
\begin{eqnarray*}
a(f) = 
\int_{{\mathbb R}^{3}}a(k)f(k)^{*}d^{3}k. 
\end{eqnarray*}
We note that such symbolical notations are 
often used in physics.

\begin{remark}
\label{remark:kernel} 
Fix $k \in {\mathbb R}^{3}$ arbitrarily. 
Then, the symbolic kernel $a(k)$ of the annihilation 
operator is given by 
\begin{eqnarray}
\left( a(k)\Psi\right)^{(n)}(k_{1}, \cdots, k_{n}) 
:= \sqrt{n+1}\Psi^{(n+1)}(k, k_{1}, \cdots, k_{n})
\label{eq:kernel-annihilation}
\end{eqnarray}
for $n = 0, 1, 2, \cdots$. 
We note that $a(k)$ is well-defined as an operator 
for $\Psi \in D_{\mathcal S} := \left\{ 
\Psi = \Psi^{(0)}\oplus\cdots\oplus 
\Psi^{(n)}\oplus\cdots \in {\mathcal F}_{0} 
\, |\, \Psi^{(n)} \in 
{\mathcal S}({\mathbb R}^{3}), n \in  {\mathbb N}\right\}$, 
where ${\mathcal S}({\mathbb R}^{3})$ is the set of all 
functions in the Schwartz class. 
The kernel $a(k)$ is defined pointwise 
by (\ref{eq:kernel-annihilation}), 
so that a certain kind of continuity is required 
for $\Psi$.  
See, for example, \cite[\S 2.2]{ammari} 
and \cite[\S 8-3]{arai-text}. 
It is well known that $a(k)^{*}$ is not densely defined 
\cite[\S X.7]{rs2}; indeed, $a(k)^{*}$ is trivial 
\cite[Proposition 8.2]{arai-text}, 
i.e., $D(a(k)^{*}) = \left\{ 0\right\}$, 
so that $a(k)$ is {\it not closable} by 
\cite[Theorem VIII.1(b)]{rs1}. 
\end{remark}

The Hilbert space in which the Hamiltonian of Nelson's model 
acts is defined by ${\mathcal H} := L^{2}({\mathbb R}^{3})
\otimes {\mathcal F}$. 
In order to define the interaction Hamiltonian 
$H_{{\mathrm I},\kappa}$ 
of Nelson's model, we use the fact 
that ${\mathcal H}$ is unitarily 
equivalent to the constant fiber 
direct integral $L^{2}({\mathbb R}^{3}, 
d^{3}x ; {\mathcal F})$, 
i.e., 
$${\mathcal H} \equiv L^{2}({\mathbb R}^{3})\otimes {\mathcal F} 
\cong  L^{2}({\mathbb R}^{3}, d^{3}x ; {\mathcal F}) 
\equiv \int^{\oplus}_{{\mathbb R}^{3}}{\mathcal F}d^{3}x,$$ 
see \cite[\S 13]{arai-text}. 
Throughout this paper, we identify ${\mathcal H}$ 
with the constant fiber direct integral, 
i.e., 
\begin{eqnarray}
{\mathcal H} = \int_{{\mathbb R}^{3}}^{\oplus}
{\mathcal F}d^{3}x. 
\label{eq:identification}
\end{eqnarray}

We set 
\begin{eqnarray*}
\lambda_{\kappa,x}(k) := 
\frac{\chi_{\kappa}(k)}{\sqrt{2\omega(k)}}\, 
e^{-ikx},\qquad 
\forall k, x \in {\mathbb R}^{3};\,\,\, 
\forall\kappa \ge 0, 
\end{eqnarray*}
where $\chi_{\kappa}(k) := (2\pi)^{-3/2}$ if  
$\kappa \le |k| \le \Lambda$; 
$:= 0$ if $|k| < \kappa$ or $\Lambda < |k|$ 
for positive constants $\kappa$ and $\Lambda$. 
Physically, $\kappa$ and $\Lambda$ mean 
an infrared cutoff and an ultraviolet cutoff, respectively. 
We fix $\Lambda$ in this paper.  
Then, we can define $H_{{\mathrm I},\kappa}$ by  
\begin{eqnarray*} 
H_{{\mathrm I},\kappa} := 
\int^{\oplus}_{{\mathbb R}^{3}}\phi_{\kappa}(x)d^{3}x,  
\end{eqnarray*}
where $\phi_{\kappa}(x)$ is the cutoff Bose field 
given by  
\begin{eqnarray*} 
\phi_{\kappa}(x) = a^{\dagger}(\lambda_{\kappa,x}) 
+ a(\lambda_{\kappa,x}).  
\end{eqnarray*}
We symbolically denote $H_{{\mathrm I},\kappa}$ by   
$$H_{{\mathrm I},\kappa}  
= 
\int_{{\mathbb R}^{3}} 
\frac{\chi_{{}_{\kappa}}(k)}{\sqrt{2\omega(k)}}
\left( e^{ikx} a(k) 
 + e^{-ikx}a^{\dagger}(k)\right)d^{3}k. 
$$  
It is well known that $H_{{\mathrm I},\kappa}$ is 
a self-adjoint operator acting in ${\mathcal H}$ 
\cite[Theorem 13-5]{arai-text}.

From now on, we also {\it denote the identity operator on all 
Hilbert spaces by $I$}. 
So, for example, $I\otimes I$ is abbreviated to $I$. 
Moreover, {\it a constant operator with the form of 
$cI$ is abbreviated 
to $c$ for a constant $c$}.

The cutoff Nelson Hamiltonian 
is given by 
\begin{eqnarray}
H^{\mbox{\rm {\tiny N}}}_{\kappa} &:=& 
H_{\mathrm {at}}\otimes I  + I\otimes H_{\mathrm f} 
+ \textsl{q}H_{{\mathrm I},\kappa}, \qquad  
\label{eq:Nelson-Hamiltonian} 
0 \le \forall\kappa < \Lambda; \,\,\, 
\forall \textsl{q} \in {\mathbb R}, 
\end{eqnarray}
acting in 
${\mathcal H} \equiv  
L^{2}({\mathbb R}^{3})\otimes {\mathcal F}$.
If the infimum of the spectrum of 
$H^{\mbox{\rm {\tiny N}}}_{\kappa}$ exists, 
we call it the {\it ground state energy} of 
$H^{\mbox{\rm {\tiny N}}}_{\kappa}$. 
Namely, the ground state energy $E^{\mbox{\tiny N}}_{\kappa}$ 
of $H^{\mbox{\rm {\tiny N}}}_{\kappa}$ is defined by 
$$
E^{\mbox{\tiny N}}_{\kappa} 
:= \inf\sigma( H^{\mbox{\tiny N}}_{\kappa}).
$$
We say that $H^{\mbox{\rm {\tiny N}}}_{\kappa}$ 
{\it has a ground state} if 
$E^{\mbox{\tiny N}}_{\kappa}$ is an eigenvalue 
of $H^{\mbox{\rm {\tiny N}}}_{\kappa}$.
In this case, every eigenvector with the 
eigenvalue $E^{\mbox{\tiny N}}_{\kappa}$ 
is called a {\it ground state}. 
Namely, the ground state $\psi_{\kappa}$ 
satisfies 
$H^{\mbox{\rm {\tiny N}}}_{\kappa}\psi_{\kappa} 
= E^{\mbox{\tiny N}}_{\kappa}\psi_{\kappa}$.    
The boson in the ground state $\psi_{\kappa}$ 
is called {\it soft boson} in this paper. 
We set 
$$H_{\mbox{\tiny N}} := H^{\mbox{\tiny N}}_{0} \equiv 
H^{\mbox{\tiny N}}_{\kappa}\lceil_{\kappa=0}$$ 
and denote the ground state energy of 
$H^{\mbox{\tiny N}}_{\kappa}$ 
and $H_{\mbox{\tiny N}}$ 
by $E^{\mbox{\tiny N}}_{\kappa}$ 
and $E_{\mbox{\tiny N}}$, respectively, i.e., 
$$E_{\mbox{\tiny N}} 
:= \inf\sigma(H_{\mbox{\tiny N}}).$$ 
Then, we have  
$$
E^{\mbox{\tiny N}}_{\kappa} \le 
\langle \psi_{\mathrm {at}}\otimes\Omega_{0}\, ,\, 
H^{\mbox{\tiny N}}_{\kappa}\psi_{\mathrm {at}}
\otimes \Omega_{0}\rangle_{\mathcal H} = 
E_{\mathrm {at}},
$$
where $\langle\quad,\quad\rangle_{\mathcal H}$ 
is the standard inner product of ${\mathcal H}$. 
We define a non-negative Hamiltonian by 
\begin{eqnarray*}
H_{\mathrm 0} := ({H_{\mathrm {at}}} 
- {E_{\mathrm {at}}})\otimes I 
+ I\otimes H_{\mathrm f}.   
\end{eqnarray*}
Then, there exist $C_{\Lambda}^{(1)}, 
C_{\Lambda}^{(2)} > 0$ such that 
$$\| H_{{\mathrm I},\kappa}\psi\|_{\mathcal H} 
\le 
C_{\Lambda}^{(1)}
\| (H_{0} + I)\psi\|_{\mathcal H} 
+ 
C_{\Lambda}^{(2)}
\| \psi\|_{\mathcal H}$$ 
for every $\psi \in  D(H_{0})$, 
which is proved in (\ref{eq:4-18-4}) below.
Combining this with a Kato-Rellich type 
argument and the variational characterization of 
eigenvalues (see, e.g., 
\cite[Theorems 13-10\, \& 13-23]{arai-text}), 
we obtain the following proposition immediately:  

\begin{proposition} 
\label{proposition:nelson-sa}
$H^{\mbox{\rm {\tiny N}}}_{\kappa}$, 
$0 \le \kappa \le \Lambda$, is self-adjoint with 
$D(H^{\mbox{\rm {\tiny N}}}_{\kappa}) 
= D(H_{\mathrm 0}) \equiv D(H_{\mathrm {at}}\otimes I)
\cap D(I\otimes H_{\mathrm f})$. 
$H^{\mbox{\rm {\tiny N}}}_{\kappa}$, $0 \le \kappa \le \Lambda$, 
is bounded from below for arbitrary 
values of $\textsl{q}$. 
In particular, 
\begin{eqnarray*}
E_{\mathrm {at}} 
- \textsl{q}^{2}\|\lambda_{\kappa,0}\|_{L^{2}}^{2} 
\le E^{\mbox{\rm {\tiny N}}}_{\kappa} 
\le E_{\mathrm {at}}. 
\end{eqnarray*}
Moreover, $H^{\mbox{\rm {\tiny N}}}_{\kappa}$, 
$0 \le \kappa \le \Lambda$, 
is essentially self-adjoint on every core for 
$H_{0}$. 
\end{proposition}

It follows from $\omega(k) = |k|$ that 
in the case $\kappa = 0$ Nelson's Hamiltonian 
$H_{\mbox{\tiny N}} \equiv  H^{\mbox{\tiny N}}_{0} 
= H^{\mbox{\tiny N}}_{\kappa}\lceil_{\kappa=0}$ 
has the singularity at 
$k = 0$ such that  
$$
\lim_{|k|\to 0}\frac{\lambda_{0,x}(k)}{\omega(k)} = \infty 
\quad\mbox{and}\quad  
\frac{\lambda_{0,x}}{\omega} \notin L^{2}({\mathbb R}^{3}). 
$$ 
On the other hand, we have 
$\lambda_{\kappa,x}/\omega \in L^{2}({\mathbb R}^{3})$ 
in the case $\kappa > 0$. 
The former condition is called {\it infrared singularity} 
(IRS) condition in \cite{ah2} (see also \cite[(3.5)]{ahh}), 
the latter {\it infrared regularity} condition.

Denote the number operator of bosons 
by $N_{\mathrm f}$, which is defined 
as the second quantization of the identity 
operator $I$, i.e., 
\begin{eqnarray}
N_{\mathrm f} := d\Gamma(I). 
\label{eq:number-operator-rigorous}
\end{eqnarray}
Symbolically, 
\begin{eqnarray*} 
N_{\mathrm f} = \int_{{\mathbb R}^{3}} a^{\dagger}(k)a(k)d^{3}k. 
\end{eqnarray*} 

In \cite[Theorem 3.2]{ahh} the absence theorem is 
described in terms of the total number of soft bosons 
forming the cloud in which the Schr\"{o}dinger 
particle is dressed. 
Namely, the statement was that ground state is 
absent from $D(I\otimes N_{\mathrm {f}}^{1/2})$. 
Our theorem is characterized 
by the spatial localization of the ground state. 
Namely,   

\begin{theorem}[absence of ground states from $D(x^{2}\otimes I)$ 
for $\kappa = 0$]
\label{theorem:2} 
Assume (A).  
For every $\textsl{q}$ with $\textsl{q} \ne 0$, 
$H_{\mbox{\rm {\tiny N}}} 
= {H^{\mbox{\rm {\tiny N}}}_{0}}$ has no ground 
state in $D(x^{2}\otimes I)$.
\end{theorem}

This theorem indirectly says that uncertainty of 
the position in ground state is infinite. 
Namely, for the ground state $\psi_{\kappa}$ 
with $\|\psi_{\kappa}\|_{\mathcal H} = 1$ we have 
symbolically   
\begin{eqnarray}
(\Delta x)_{\mathrm {gs}} := 
\langle\psi_{\kappa}\, ,\, 
(x\otimes I - \langle x\rangle_{\mathrm {gs}})^{2}
\psi_{\kappa}\rangle_{\mathcal H}^{1/2} = \infty, 
\label{eq:uncertainty-position}
\end{eqnarray}
where $\langle x\rangle_{\mathrm {gs}}$ is 
the expectation vector of the position in 
the ground state,  
\begin{eqnarray*}
\langle x\rangle_{\mathrm {gs}} 
:= \langle\psi_{\kappa}\, ,\, 
x\otimes I\psi_{\kappa}\rangle_{\mathcal H} 
\in  {\mathbb R}^{3}. 
\end{eqnarray*}

\begin{theorem}[non-existence of any ground state for $\kappa = 0$]
\label{crl:2} 
Let $V$ be in class (C1) or (C2). 
Then, for every $\textsl{q}$ with $\textsl{q} \ne 0$, 
$H_{\mbox{\rm {\tiny N}}} 
= {H^{\mbox{\rm {\tiny N}}}_{0}}$ has no ground 
state in ${\mathcal H}$.
\end{theorem}

Without loss of generality, we have only to consider 
a normalized ground state. Thus, {\it we always treat 
the normalized ground state throughout this paper}. 

\begin{theorem}[soft-boson divergence]
\label{theorem:1} 
Assume (A) and that there exists a constant 
$\textsl{q}_{{}_{0}}$ such that 
$H_{\kappa}^{\mbox{\rm {\tiny N}}}$ 
has a (normalized) ground state $\psi_{\kappa}$ 
for every $\kappa$ with $0 < \kappa < \Lambda$ and 
$\textsl{q}$
with $|\textsl{q}| < \textsl{q}_{{}_{0}}$. 
If $\psi_{\kappa} \in  D(x^{2}\otimes I)$, 
then  
\begin{eqnarray}  
\qquad 
&{}& 
\left\{
\frac{\textsl{q}^{2}}{8\pi^{2}}\left( 
\log\frac{\Lambda}{\kappa}\right) 
- 
\frac{\textsl{q}^{2}}{8\pi^{2}}\Lambda^{2}
\| |x|\otimes I\psi_{\kappa}
\|_{\mathcal H}^{2}
 \right\} 
\label{eq:ine1} \\ 
\nonumber  
&{}& \qquad 
\le \,\,\, 
\langle \psi_{\kappa}\, ,\, 
I\otimes N_{\mathrm f}\psi_{\kappa}\rangle_{\mathcal H} \\ 
\nonumber 
&{}& \qquad\qquad  
\le 
\left\{
\frac{\textsl{q}^{2}}{2\pi^{2}}\left( 
\log\frac{\Lambda}{\kappa}\right) 
+ 
\frac{\textsl{q}^{2}}{4\pi^{2}}\Lambda^{2}
\| |x|\otimes I\psi_{\kappa}
\|_{\mathcal H}^{2}
\right\}.  
\end{eqnarray}
\end{theorem}

For the case where $V$ is in class (C2), 
we define a positive constant 
$\textsl{q}_{{}_{\Lambda}}$ by 
\begin{eqnarray*}
\Sigma - {E_{\mathrm {at}}} =  
\frac{\textsl{q}_{{}_{\Lambda}}^{2}}{4(2\pi)^{3}}
\int_{|k| \le \Lambda} 
\frac{|k|}{|k| + k^{2}/2}d^{3}k,  
\end{eqnarray*}
where $\Sigma := \inf\sigma_{\mathrm {ess}}
(H_{\mathrm {at}})$. 
We set 
$\textsl{q}_{{}_{\Lambda}} = \infty$ for 
the case where $V$ is in class (C1) 
because $\Sigma = \infty$ in this case. 
Note that $\textsl{q}_{{}_{\Lambda}}$ is independent of 
$\kappa$. 
By \cite[Proposition III.3]{gerard} and  
\cite[Theorem 1]{spohn} and noting 
$$\frac{\Sigma - E_{\mathrm {at}}}{\displaystyle \,\,\,
\frac{1}{2}\int_{{\mathbb R}^{3}} |\lambda_{\kappa,x}(k)|^{2}
k^{2}\left(\omega(k) + k^{2}/2\right)^{-1}d^{3}k
\,\,\,} \ge \textsl{q}_{{}_{\Lambda}}^{2},$$ 
we have the following proposition. 
\begin{proposition}
\label{proposition:gs-existence}
Let us fix $\Lambda > 0$.   
$H^{\mbox{\rm {\tiny N}}}_{\kappa}$ 
has a unique ground 
state $\psi_{\kappa}$ for every 
$\kappa, \textsl{q}$ with $0 < \kappa < \Lambda$ 
and $|\textsl{q}| < \textsl{q}_{{}_{\Lambda}}$,  
provided that $V$ is in class (C1) or (C2). 
\end{proposition}

For these ground states $\psi_{\kappa}$, 
$0 < \kappa < \Lambda$, we have the following: 

\begin{theorem}[soft-boson divergence]
\label{crl:1} Let $V$ be in (C1) or (C2). 
Then, for the ground states $\psi_{\kappa}$ of 
$H^{\mbox{\rm {\tiny N}}}_{\kappa}$, $0 < \kappa 
< \Lambda$, (\ref{eq:ine1}) holds. 
Moreover, $\sup_{0<\kappa<\Lambda}$
$\| |x|\otimes I\psi_{\kappa}\|_{\mathcal H} < \infty$ 
and  
\begin{eqnarray*}  
\qquad 
&{}& 
\left\{
\frac{\textsl{q}^{2}}{8\pi^{2}}\left( 
\log\frac{\Lambda}{\kappa}\right) 
- 
\frac{\textsl{q}^{2}}{8\pi^{2}}\Lambda^{2}
\sup_{0<\kappa<\Lambda}\| |x|\otimes I\psi_{\kappa}
\|_{\mathcal H}^{2}
 \right\}  \\ 
\nonumber  
&{}& \qquad 
\le \,\,\, 
\langle \psi_{\kappa}\, ,\, 
I\otimes N_{\mathrm f}\psi_{\kappa}\rangle_{\mathcal H} \\ 
\nonumber 
&{}& \qquad\qquad  
\le 
\left\{
\frac{\textsl{q}^{2}}{2\pi^{2}}\left( 
\log\frac{\Lambda}{\kappa}\right) 
+ 
\frac{\textsl{q}^{2}}{4\pi^{2}}\Lambda^{2}
\sup_{0<\kappa<\Lambda}\| |x|\otimes I\psi_{\kappa}
\|_{\mathcal H}^{2}
\right\}.  
\end{eqnarray*}
\end{theorem}

We prove Theorem \ref{theorem:2} and Theorem \ref{theorem:1} 
in \S \ref{section:ABG} and \S \ref{section:EESPB}, 
respectively. 
Combining these theorems with the fact on uncertainty 
argued in \S \ref{section:FNPGS}, Theorems \ref{crl:2} 
and \ref{crl:1} 
are also proved in \S \ref{section:FNPGS}.

\section{An identity from the operator-theoretical 
pull-through formula} 
\label{section:RPTF} 

Let us fix $0 \le \kappa < \Lambda$, 
and we suppose that $H^{\mbox{\tiny N}}_{\kappa}$ 
has a ground state $\psi_{\kappa}$ 
throughout this section. 
As declared before Theorem \ref{theorem:1}, 
for simplicity we normalized $\psi_{\kappa}$ throughout. 
By using the kernel version of CCR, $[ a(k)\, ,\, a^{\dagger}(k')] 
= \delta(k-k')$, we symbolically obtain 
the pull-through formula on the ground state $\psi_{\kappa}$,  
\begin{eqnarray} 
I\otimes a(k)\psi_{\kappa} 
= 
- \textsl{q}\frac{\chi_{\kappa}(k)}{\sqrt{2\omega(k)}}
\left( H^{\mbox{\tiny N}}_{\kappa} 
- E^{\mbox{\tiny N}}_{\kappa} + \omega(k)\right)^{-1} 
e^{-ikx}\otimes I\psi_{\kappa}.  
\label{eq:pull-through-formula} 
\end{eqnarray}
However, since the domain of $a(k)$ is so narrow 
that $a(k)$ is not closable 
as remarked in Remark \ref{remark:kernel}, 
(\ref{eq:pull-through-formula}) itself should {\it not} 
be regarded as an operator equality on ground states.    
It should be regarded as an equality on 
$L^{2}_{\mathrm {loc}} 
({\mathbb R}^{3};{\mathcal H})$  
as Derezi\'{n}ski and G\'{e}rard did in \cite[Theorem 2.5]{dg}. 
The purposes of this section is to prove 
the operator-theoretical pull-through formula 
on the ground state and derive a useful decomposition 
for Nelson's model from it. 
To author's best knowledge,  
the proof in this paper is the first 
for the pull-through formula in operator theory 
and the operator-theoretical version of this formula 
has another development in operator theory 
of IR catastrophe (cf. \cite{ahh2,hirokawa-IR}). 
 
Before we state our desired proposition, 
we note the following lemma. 

\begin{lemma}
\label{lemma:Liouville}
For $f \in  L^{2}({\mathbb R}^{3})$ and 
$t \in  {\mathbb R}$, set 
\begin{eqnarray*}
a_{t}(f) := e^{itH^{\mbox{\tiny N}}_{\kappa}}
\left( I\otimes 
a(e^{-i\omega t}f)\right)
e^{-itH^{\mbox{\tiny N}}_{\kappa}}.
\end{eqnarray*}
If $\omega f, 
f/\sqrt{\omega} \in  L^{2}({\mathbb R}^{3})$, then 
\begin{eqnarray} 
\qquad\frac{d}{dt}a_{t}(f)\psi 
= 
-\,i\textsl{q}e^{itH_{\kappa}^{\mbox{\tiny N}}}
\left\{\left(\int_{{\mathbb R}^{3}}f(k)^{*}
e^{it\omega(k)}\lambda_{\kappa,x}(k)d^{3}k\right)
\otimes I\right\}
e^{- itH_{\kappa}^{\mbox{\tiny N}}}
\psi 
\label{eq:Liouville}
\end{eqnarray}
for every 
$\psi \in  D((H_{\kappa}^{\mbox{\tiny N}})^{2})$.  
\end{lemma}

\proof 
In the same way as in \cite[Theorem 4.1]{km}, 
we can prove that 
\begin{eqnarray*} 
\frac{d}{dt}a_{t}(f)\psi 
&=& 
ie^{itH_{\kappa}^{\mbox{\tiny N}}} 
\left[ \textsl{q}H_{{\mathrm I},\kappa}\, ,\, 
I\otimes a(e^{-i\omega t}f)
\right] 
e^{-itH_{\kappa}^{\mbox{\tiny N}}}\psi
\end{eqnarray*}
for every 
$\psi \in  D((H_{\kappa}^{\mbox{\tiny N}})^{2})$. 
We obtain (\ref{eq:Liouville}) from this 
equation directly.
\qed

\begin{proposition}[pull-through formula on ground states] 
\label{proposition:pull-through-formula}
Fix $\kappa$ with $0 \le \kappa < \Lambda$. 
Assume (A) and suppose that 
$H_{\kappa}^{\mbox{\rm {\tiny N}}}$ 
has a ground state $\psi_{\kappa}$. 
If $\psi_{\kappa} \in  
D(x^{2}\otimes I)$, 
then for all $f \in  C_{0}^{\infty}({\mathbb R}^{3}
\setminus \left\{ 0\right\})$, 
\begin{eqnarray} 
&{}& I\otimes a(f)\psi_{\kappa} 
\label{eq:pull-through-formula-rigorous} \\ 
\nonumber 
&{}& 
= 
- \textsl{q} 
\int_{{\mathbb R}^{3}}
f(k)^{*}
\frac{\chi_{\kappa}(k)}{\sqrt{2\omega(k)}}
\left( H^{\mbox{\rm {\tiny N}}}_{\kappa} 
- E^{\mbox{\rm {\tiny N}}}_{\kappa} + \omega(k)\right)^{-1} 
\left(e^{-ikx}\otimes I\right)\psi_{\kappa}d^{3}k.  
\end{eqnarray}
\end{proposition}

\proof
Let $f \in  C_{0}^{\infty}({\mathbb R}^{3}
\setminus \left\{ 0\right\})$. 
Then, there exists $d_{f} > 0$ such that 
$\left\{ k \in  {\mathbb R}^{3}\, |\, |k| < d_{f}\right\}$ 
$\subset$ ${\mathbb R}^{3}\setminus {\mathrm {supp}}f$, 
which implies ${\mathrm {supp}}f \subset \left\{ k \in  
{\mathbb R}^{3} |\, |k| > d_{f}/2\right\}$. 
Set $\Omega_{\kappa,\Lambda}^{\mathrm {int}}$ 
$:=$ $\left\{\right.$ $k \in  {\mathbb R}^{3}$\, 
$|$\, $\kappa < |k| < \Lambda$$\left.\right\}$ and 
$\Omega_{\kappa,\Lambda}^{\mathrm {ext}}$ 
$:=$ $\left\{ k \in  {\mathbb R}^{3}\, |\,
\right.$ 
$0 < |k| < \kappa\,\,\,\mbox{or}$ 
$\left. \Lambda < |k|\right\}$. 
Since $L^{2}({\mathbb R}^{3}) \cong 
L^{2}(\Omega_{\kappa,\Lambda}^{\mathrm {int}})
\oplus L^{2}(\Omega_{\kappa,\Lambda}^{\mathrm {ext}})$, 
we identify $L^{2}({\mathbb R}^{3})$ with 
$L^{2}(\Omega_{\kappa,\Lambda}^{\mathrm {int}})
\oplus L^{2}(\Omega_{\kappa,\Lambda}^{\mathrm {ext}})$ 
in this proof. 
There exists $f^{\sharp} 
\in  L^{2}(\Omega_{\kappa,\Lambda}^{\sharp})$, 
$\sharp = {\mathrm {int}}, 
{\mathrm {ext}}$, such that 
$L^{2}({\mathbb R}^{3}) \ni f = f^{\mathrm {int}}
\oplus f^{\mathrm {ext}} \in 
L^{2}(\Omega_{\kappa,\Lambda}^{\mathrm {int}})
\oplus L^{2}(\Omega_{\kappa,\Lambda}^{\mathrm {ext}})$. 
For $f^{\sharp}$, there exists a sequence $f_{\nu}^{\sharp} 
\in  C_{0}^{\infty}(\Omega_{\kappa,\Lambda}^{\sharp})$, 
$\nu \in  {\mathbb N}$, such that 
$f^{\sharp}_{\nu} \to f^{\sharp}$ in 
$L^{2}(\Omega_{\kappa,\Lambda}^{\sharp})$ 
as $\nu\to\infty$ and 
${\mathrm {supp}}(f_{\nu}^{\mathrm {int}}\oplus 
f_{\nu}^{\mathrm {ext}}) \subset \left\{ k \in  
{\mathbb R}^{3} |\, |k| > d_{f}/2\right\}$ for 
each $\nu$. 
For simplicity, we denote $f_{\nu}^{\mathrm {int}}\oplus 
f_{\nu}^{\mathrm {ext}}$ by $f_{\nu}$, i.e., 
$f_{\nu} := f_{\nu}^{\mathrm {int}}\oplus 
f_{\nu}^{\mathrm {ext}}$. 

For every $\psi \in  D((H_{\kappa}^{\mbox{\tiny N}})^{2})$, 
$t \in  {\mathbb R}$, and the above $f_{\nu}$,  
we have  
\begin{eqnarray*}
a_{t}(f_{\nu})\psi &=&  
I\otimes a(f_{\nu})\psi \\ 
&{}& 
-i\textsl{q}
\int_{0}^{t}e^{isH^{\mbox{\tiny N}}_{\kappa}}
\left\{
\left( 
\int_{{\mathbb R}^{3}}
f_{\nu}(k)^{*}e^{is\omega(k)}\lambda_{\kappa,x}(k)d^{3}k
\right)\otimes I\right\}
e^{-isH^{\mbox{\tiny N}}_{\kappa}}\psi ds
\end{eqnarray*}
by Lemma \ref{lemma:Liouville}. 
Here, we note that ${\mathrm {supp}}
(f_{\nu}^{*}\lambda_{\kappa,x}) = 
{\mathrm {supp}}((f_{\nu}^{\mbox{\rm {\tiny int}}})^{*}
\lambda_{\kappa,x})$. 
Since $f_{\nu}^{*}\lambda_{\kappa,x} 
\in  C_{0}^{\infty}(\Omega_{\kappa,\Lambda}^{\mathrm {int}})$, 
we obtain by partial integration as in \cite[Lemma 4.3]{ahh} 
that 
\begin{eqnarray*}
\int_{{\mathbb R}^{3}}f_{\nu}(k)^{*}
e^{it\omega(k)}\lambda_{\kappa,x}(k)d^{3}k 
= - \frac{1}{t^{2}}
\int_{{\mathbb R}^{3}}g(k)e^{it\omega(k)}d^{3}k, 
\end{eqnarray*}
where for $n, m = 1, 2, 3$,
\begin{eqnarray*} 
g(k) &=&  
\partial_{m}\left\{\frac{1}{\partial_{m}\omega(k)}
\partial_{n}\left(\frac{1}{\partial_{n}\omega(k)}
\lambda_{\kappa,x}(k)
f_{\nu}(k)^{*}\right)\right\} \\ 
&{}& \qquad\, 
=  
\partial_{m}\left\{\frac{1}{\partial_{m}\omega(k)}
\partial_{n}\left(e^{-ikx}\frac{1}{\partial_{n}\omega(k)}
\lambda_{\kappa,0}(k)
f_{\nu}(k)^{*}\right)\right\},    
\end{eqnarray*}
with $\partial_{n} := 
\partial/\partial k_{n}$. 
Concerning $\partial_{n}\lambda_{\kappa,x}$ and 
$\partial_{m}\partial_{n}\lambda_{\kappa,x}$ 
in the above expression of $g(k)$, 
we can directly estimate them in the 
following because the function of $x$ appearing 
in $\lambda_{\kappa,x}$ is only $e^{-ikx}$. 
There exists $C_{\Lambda,\nu} > 0$, which is independent 
of $\kappa, x$, such that 
$|\partial_{n}\lambda_{\kappa,x}(k)| \le 
C_{\Lambda,\nu}(1+|x|)$ and 
$|\partial_{m}\partial_{n}\lambda_{\kappa,x}(k)| \le 
C_{\Lambda,\nu}(1+|x|^{2})$ for every $k$ with 
$\kappa < |k| < \Lambda$. 
Thus, we have $g \in  L({\mathbb R}^{3})$ 
and we can show that 
$a_{\pm}(f_{\nu})\psi$ $=$ $s\mbox{-}\lim_{t\to\pm\infty}$
$a_{t}(f_{\nu})\psi$ 
exists for all 
$\psi \in D({H_{\kappa}^{\mbox{\tiny N}}}^{2})\cap 
D(x^{2}\otimes I)$ in the same way as in 
\cite[Lemma 4.3]{ahh}. 
So, we have the following equality
\begin{eqnarray*}
a_{\pm}(f_{\nu})\psi &=&  
I\otimes a(f_{\nu})\psi \\ 
&{}& 
-i\textsl{q}
\int_{0}^{\pm\infty}e^{itH^{\mbox{\tiny N}}_{\kappa}}
\left\{\left( 
\int_{{\mathbb R}^{3}}
f_{\nu}(k)^{*}e^{it\omega(k)}\lambda_{\kappa,x}(k)d^{3}k
\right)\otimes I \right\}
e^{-itH^{\mbox{\tiny N}}_{\kappa}}\psi dt.
\end{eqnarray*}
Also see \cite[Theorem 1 and (6)]{hk} and  
\cite[Theorem 5.1]{km}.  
Moreover, using the absolute continuity of $\omega(k)$ 
and the Riemann-Lebesgue theorem, 
we have $a_{\pm}(f_{\nu})\psi_{\kappa} = 0$. 
By using these facts and $e^{-itH^{\mbox{\tiny N}}_{\kappa}}
\psi_{\kappa} = e^{-itE^{\mbox{\tiny N}}_{\kappa}}\psi_{\kappa}$, 
we have 
\begin{eqnarray*}
&{}& 
I\otimes a(f_{\nu})\psi_{\kappa}     \\
\nonumber 
&=& 
i\textsl{q}\int_{0}^{\infty} 
e^{it(H^{\mbox{\tiny N}}_{\kappa} - 
E^{\mbox{\tiny N}}_{\kappa})} 
\left( 
\int_{{\mathbb R}^{3}}f_{\nu}(k)^{*}
e^{it\omega(k)}\lambda_{\kappa, x}(k)d^{3}k 
\right)\otimes I\psi_{\kappa}dt.
\end{eqnarray*} 
So, by Fubini's theorem and Lebesgue's dominated convergence 
theorem, we have for every $\phi \in  
D(H_{\kappa}^{\mbox{\rm {\tiny N}}})$ 
\begin{eqnarray}
\qquad&{}& 
\langle\phi, 
I\otimes a(f_{\nu})\psi_{\kappa}\rangle_{\mathcal H} 
\label{eq:pt1} \\ 
\nonumber 
&=& 
i\textsl{q}\lim_{\varepsilon\downarrow 0}
\int_{0}^{\infty}e^{-t\varepsilon}
\left( 
\int_{{\mathbb R}^{3}}
f_{\nu}(k)^{*}
\langle\phi\, ,\, e^{it(H_{\kappa}^{\mbox{\rm {\tiny N}}} 
- E_{\kappa}^{\mbox{\rm {\tiny N}}} + \omega(k))}
\lambda_{\kappa,x}\otimes I\psi_{\kappa}\rangle_{\mathcal H}
d^{3}k\right)dt \\ 
\nonumber 
&=& 
i\textsl{q}\lim_{\varepsilon\downarrow 0}
\int_{{\mathbb R}^{3}}
f_{\nu}(k)^{*}
\bigg< 
\int_{0}^{\infty}
e^{-it(H_{\kappa}^{\mbox{\rm {\tiny N}}} 
- E_{\kappa}^{\mbox{\rm {\tiny N}}} + \omega(k) - i\varepsilon)}
\phi dt\, ,\, 
\lambda_{\kappa,x}\otimes I\psi_{\kappa}
\bigg>_{\mathcal H}
d^{3}k  \\ 
\nonumber 
&=& 
i\textsl{q}\lim_{\varepsilon\downarrow 0}
\int_{{\mathbb R}^{3}}
f_{\nu}(k)^{*}
\langle 
-i(H_{\kappa}^{\mbox{\rm {\tiny N}}} 
- E_{\kappa}^{\mbox{\rm {\tiny N}}} + \omega(k) - i\varepsilon)^{-1}
\phi \, ,\, 
\lambda_{\kappa,x}\otimes I\psi_{\kappa}\rangle_{\mathcal H}
d^{3}k \\ 
\nonumber 
&=& 
\bigg< 
\phi \, ,\, 
-\textsl{q}
\int_{{\mathbb R}^{3}}
f_{\nu}(k)^{*}
(H_{\kappa}^{\mbox{\rm {\tiny N}}} 
- E_{\kappa}^{\mbox{\rm {\tiny N}}} + \omega(k))^{-1} 
\lambda_{\kappa,x}\otimes I\psi_{\kappa} d^{3}k
\bigg>_{\mathcal H},  
\end{eqnarray}
where we used Fubini's theorem in the $2$nd equality noting 
\begin{eqnarray*}
&{}& \biggl|
e^{-t\varepsilon}f_{\nu}(k)^{*}
\langle e^{-it(H_{\kappa}^{\mbox{\rm {\tiny N}}} 
- E_{\kappa}^{\mbox{\rm {\tiny N}}} + \omega(k))}\phi\, ,\, 
\lambda_{\kappa,x}\otimes I\psi_{\kappa}
\rangle_{\mathcal H} 
\biggr| \\ 
&\le&  
e^{-t\varepsilon}|f_{\nu}(k)|\, |\lambda_{\kappa,x}(k)|\, 
\|\phi\|_{\mathcal H}, 
\end{eqnarray*} 
and 
we calculated the integral over $0 < t < \infty$ 
in the $3$rd equality using 
\begin{eqnarray*}
&{}& 
\lim_{T\to\infty}
i(H_{\kappa}^{\mbox{\rm {\tiny N}}} 
- E_{\kappa}^{\mbox{\rm {\tiny N}}} + \omega(k) - i\varepsilon)^{-1}
e^{-iT(H_{\kappa}^{\mbox{\rm {\tiny N}}} 
- E_{\kappa}^{\mbox{\rm {\tiny N}}} + \omega(k) - i\varepsilon)}\phi \\ 
&=&  
\lim_{T\to\infty}
e^{-T\varepsilon}
i(H_{\kappa}^{\mbox{\rm {\tiny N}}} 
- E_{\kappa}^{\mbox{\rm {\tiny N}}} + \omega(k) - i\varepsilon)^{-1}
e^{-iT(H_{\kappa}^{\mbox{\rm {\tiny N}}} 
- E_{\kappa}^{\mbox{\rm {\tiny N}}} + \omega(k))}\phi 
= 0.  
\end{eqnarray*} 
Therefore, (\ref{eq:pull-through-formula-rigorous}) 
for $f_{\nu}$ follows from (\ref{eq:pt1}). 

If $k \in  {\mathrm {supp}}f \cup 
\left(\bigcup_{\nu \ge \nu_{0}} 
{\mathrm {supp}}f_{\nu}\right)$, then 
$|k|^{-1} < 2/d_{f}$. 
Hence it follows that  
$\|f_{\nu}/\sqrt{\omega} - f/\sqrt{\omega}
\|_{L^{2}}^{2} 
\le 2d_{f}^{-1}
\|f_{\nu} - f
\|_{L^{2}}^{2} 
= 2d_{f}^{-1}(\|f_{\nu}^{\mathrm {int}} - 
f^{\mathrm {int}}\|_{L^{2}(
\Omega_{\kappa,\Lambda}^{\mathrm {int}})}^{2} 
+
\|f_{\nu}^{\mathrm {ext}} - 
f^{\mathrm {ext}}\|_{L^{2}(
\Omega_{\kappa,\Lambda}^{\mathrm {ext}})}^{2} 
) 
$ 
for $\nu \ge \nu_{0}$. 
Therefore, we obtain 
\begin{eqnarray}
f_{\nu}/\omega^{j/2} \longrightarrow 
f/\omega^{j/2}
\label{eq:new-12-*}
\end{eqnarray}
in $L^{2}({\mathbb R}^{3})$ as 
$\nu\to\infty$ for $j = 0, 1$. 
Since $\psi_{\kappa} \in  D(H_{0}^{1/2})$ 
by Proposition \ref{proposition:nelson-sa}, 
the fundamental inequality 
$\| I\otimes a(f_{\nu})\psi_{\kappa} - 
I\otimes a(f)\psi_{\kappa}\|_{\mathcal H} 
\le \| (f_{\nu} - f)/\sqrt{\omega}\|_{L^{2}}
\|I\otimes H_{0}^{1/2}\psi_{\kappa}\|_{\mathcal H}$ 
holds. 
So, by (\ref{eq:new-12-*}), 
$I\otimes a(f_{\nu})\psi_{\kappa} \longrightarrow 
I\otimes a(f)\psi_{\kappa}$ as $\nu\to\infty$. 
By the Schwarz inequality, 
$\chi_{{}_{0}}\omega^{-1} \in  L^{2}({\mathbb R}^{3})$, 
and (\ref{eq:new-12-*}), 
the r.h.s of (\ref{eq:pull-through-formula-rigorous}) 
for $f_{\nu}$ converges to 
that for $f$. Therefore, 
(\ref{eq:pull-through-formula-rigorous}) holds 
for $f \in  C_{0}^{\infty}({\mathbb R}^{3}
\setminus \left\{ 0\right\})$. 
\qed

In (\ref{eq:pull-through-formula}), 
we employ the following decomposition of the plain wave 
$e^{-ikx}$ into the dipole-approximated term 
$e^{-ik0} = 1$ and the error term $e^{-ikx} - 1$, 
i.e., 
\begin{eqnarray}
e^{-ikx} = 1 + (e^{-ikx} - 1), 
\label{eq:decomposition}
\end{eqnarray}
because this decomposition provides very simple 
treatment to estimate the total number of soft bosons. 
Derezi\'{n}ski and G\'{e}rard implement this way 
in $L^{2}$-theory \cite{dg}. 
We also employ this way and implement it in operator theory 
by using (\ref{eq:pull-through-formula-rigorous}).

\begin{proposition} 
\label{proposition:a-representation-simple} 
Let us fix $\kappa$ with $0 \le \kappa < \Lambda$, and 
suppose that $H^{\mbox{\rm {\tiny N}}}_{\kappa}$ 
has a ground state $\psi_{\kappa}$ and 
$\psi_{\kappa} \in  D(x^{2}\otimes I)$. 
Then, for all $f \in  C_{0}^{\infty}({\mathbb R}^{3}
\setminus \left\{ 0\right\})$, 
\begin{eqnarray}
I\otimes a(f)\psi_{\kappa} 
= 
\sum_{j=1}^{2}
\int_{{\mathbb R}^{3}}f(k)^{*}
J_{j}(k)\psi_{\kappa}d^{3}k   
\label{eq:a-representation-simple}
\end{eqnarray} 
with 
\begin{eqnarray*}
&{}& J_{1}(k) = 
- \textsl{q}\frac{\chi_{\kappa}(k)}{
\sqrt{2\omega(k)}\,\omega(k)}
I\otimes I,  \\ 
&{}& J_{2}(k) = 
- \textsl{q}\frac{\chi_{\kappa}(k)}{
\sqrt{2\omega(k)}}( H_{\kappa}^{\mbox{\rm {\tiny N}}} 
- E_{\kappa}^{\mbox{\rm {\tiny N}}} + \omega(k) )^{-1}
(e^{-ikx} - 1)\otimes I. 
\end{eqnarray*}
Then, 
\begin{eqnarray}
&{}& 
\int_{{\mathbb R}^{3}}
\| J_{1}(k)\psi_{\kappa}\|_{\mathcal H}^{2}d^{3}k 
= 
\frac{\textsl{q}^{2}}{4\pi^{2}}
\log\frac{\Lambda}{\kappa}, 
\label{eq:estimate-J1} \\ 
&{}& 
\int_{{\mathbb R}^{3}}
\| J_{2}(k)\psi_{\kappa}\|_{\mathcal H}^{2}d^{3}k 
\le 
\frac{\textsl{q}^{2}}{8\pi^{2}}
\Lambda^{2}\| |x|\otimes I\psi_{\kappa}\|_{\mathcal H}^{2}. 
\label{eq:estimate-J2} 
\end{eqnarray}
\end{proposition}

\proof
We obtain immediately (\ref{eq:a-representation-simple}) 
from (\ref{eq:pull-through-formula-rigorous}) by 
using (\ref{eq:decomposition}) and 
$( H_{\kappa}^{\mbox{\rm {\tiny N}}} 
- E_{\kappa}^{\mbox{\rm {\tiny N}}} + \omega(k) )^{-1}\psi_{\kappa} 
= \omega(k)^{-1}\psi_{\kappa}$. 
(\ref{eq:estimate-J1}) follows from a direct computation. 
By using $|e^{-ikx} - 1| \le |k||x|$, we have (\ref{eq:estimate-J2}). 
\qed

\begin{remark} 
\label{rem:a-representation} 
We note that decomposition (\ref{eq:decomposition}) 
is not always useful in proving the non-existence 
of ground state. 
We have to use another technique in a general case 
(e.g. see GSB model and some polaron models 
\cite{hirokawa-IR}). 
In fact, to treat several sorts of polarons, 
we mathematically consider more general 
dispersion relations $\omega(k)$ 
and coupling functions $\lambda_{\kappa,x}(k)$. 
For simplicity, we consider $\omega(k) = |k|^{\mu}$ and 
$\lambda_{\kappa,x}(k) 
= \chi_{\kappa}(k)|k|^{-\nu}e^{-ikx}$ now, 
where $\mu \ge 0$, $\nu \in  {\mathbb R}$, and 
$d = 1, 2, 3$. 
Then, because we do not always have 
(\ref{eq:estimate-J2}), our argument 
in \S \ref{section:ABG} 
does not work. For example, 
consider the case $\mu + 2\nu < d \le 
2\mu + 2\nu - 2$. 
For such a case, by following the idea in \cite{ahh} instead of 
(\ref{eq:decomposition}),  
we can press forward with a concrete computation 
from \cite[Lemma 5.1]{ahh} as announced in \cite{hirokawa-e}. 
For further details, see \cite{hirokawa-IR}. 
\end{remark}

\section{Absence of ground state from 
$D(x^{2}\otimes I)$ for $\kappa = 0$}
\label{section:ABG}

In \cite{ahh} we proved that any ground state 
of GSB model is absent from $D(I\otimes N_{\mathrm f}^{1/2})$. 
Here, by employing decomposition 
(\ref{eq:a-representation-simple}), 
we prove Theorem \ref{theorem:2}, 
namely, any ground state 
of $H_{\mbox{\rm {\tiny N}}} 
= {H^{\mbox{\rm {\tiny N}}}_{0}}$ 
is absent from $D(x^{2}\otimes I)$.

\qquad 

\underbar{\it Proof of Theorem \ref{theorem:2}}: 
We use reductio ad absurdum to 
prove Theorem \ref{theorem:2}. 
Suppose that $H_{\mbox{\tiny N}} := H^{\mbox{\tiny N}}_{0}$ 
has a ground state 
${\psi_{{}_{0}}}$ in $D(x^{2}\otimes I)$. 
We note we already normalized the ground state 
${\psi_{{}_{0}}}$. 
For every $\phi \in D(I\otimes N_{\mathrm f}^{1/2})$, 
define the function $F_{\phi,{\psi_{{}_{0}}}}$ by  
\begin{eqnarray} 
F_{\phi,{\psi_{{}_{0}}}}(k) 
= \sum_{j=1}^{2}
\langle \phi\, ,\, J_{j}(k){\psi_{{}_{0}}}
\rangle_{\mathcal H}. 
\label{eq:F}
\end{eqnarray}
Since $D(I\otimes N_{\mathrm {f}}^{1/2}) \subset 
D(I\otimes a^{\dagger}(f))$, 
we can define the anti-linear functional 
$T_{\phi,{\psi_{{}_{0}}}} : L^{2}({\mathbb R}^{3}) 
\to  {\mathbb C}$ by 
\begin{eqnarray*} 
T_{\phi,{\psi_{{}_{0}}}}(f) = 
\langle I\otimes a^{\dagger}(f)\phi\, ,\, {\psi_{{}_{0}}}
\rangle_{\mathcal H}, \qquad 
\forall\phi \in  
D(I\otimes N_{\mathrm f}^{1/2}).  
\end{eqnarray*}
By the fundamental inequality concerning 
$a^{\dagger}(f)$ and $N_{\mathrm f}$, 
we have 
\begin{eqnarray*}
|T_{\phi,{\psi_{{}_{0}}}}(f)| 
\le \|I\otimes (N_{\mathrm f} + 1)^{1/2}\phi\|_{\mathcal H}
\|f\|_{L^{2}}, 
\end{eqnarray*}
namely, 
$T_{\phi,{\psi_{{}_{0}}}}$ is a bounded anti-linear functional. 
So, by Riesz's lemma, there exists a unique $F \in 
L^{2}({\mathbb R}^{3})$ such that 
$T_{\phi,{\psi_{{}_{0}}}}(f) = 
\langle f\, ,\, F\rangle_{L^{2}}$ 
for every $f \in  
L^{2}({\mathbb R}^{3})$.   
We note that 
${\psi_{{}_{0}}} \in D( H_{\mbox{\tiny N}}) = 
D(H_{\mathrm 0}) \subset  D(H_{\mathrm 0}^{1/2}) \subset 
D(H_{\mathrm f}^{1/2}) \subset  D(a(g))$ 
for every $g \in  L^{2}({\mathbb R}^{3})$ 
with $g/\sqrt{\omega} \in  L^{2}({\mathbb R}^{3})$. 
By (\ref{eq:a-representation-simple}), 
we obtain 
$\langle f\, ,\, F_{\phi,{\psi_{{}_{0}}}}\rangle_{L^{2}} 
= \langle\phi\, ,\, I\otimes a(f){\psi_{{}_{0}}}
\rangle_{\mathcal H} 
= T_{\phi,{\psi_{{}_{0}}}}(f)$ 
for $f \in  C_{0}^{\infty}({\mathbb R}^{3}
\setminus \left\{ 0\right\})$. 
Thus, we have 
\begin{eqnarray*} 
F_{\phi,{\psi_{{}_{0}}}} = F \in  L^{2}({\mathbb R}^{3}), 
\qquad 
\forall\phi \in D(I\otimes N_{\mathrm f}^{1/2}). 
\end{eqnarray*}
By (\ref{eq:a-representation-simple}) and (\ref{eq:F}), 
we have  
\begin{eqnarray}
- \textsl{q}\Theta_{1}(k)\langle\phi\, ,\, \psi_{{}_{0}}
\rangle_{\mathcal H} 
= 
\langle\phi\, ,\, J_{1}(k){\psi_{{}_{0}}}
\rangle_{\mathcal H}
= 
F_{\phi,{\psi_{{}_{0}}}}(k) 
- 
\langle\phi\, ,\, J_{2}(k){\psi_{{}_{0}}}
\rangle_{\mathcal H} 
\label{eq:identity} 
\end{eqnarray}
as an $L^{2}({\mathbb R}^{3})$-function of $k$, 
where 
$$
\Theta_{1}(k) = 
\frac{\chi_{{}_{0}}(k)}{\sqrt{2\omega(k)}\,\omega(k)}. 
$$ 
So, by (\ref{eq:estimate-J1}) and (\ref{eq:estimate-J2}), 
we reach a contradiction if $\langle\phi\, ,\, \psi_{{}_{0}}
\rangle_{\mathcal H} \ne 0$. Namely, 
the left hand side of (\ref{eq:identity}) 
is not in $L^{2}({\mathbb R}^{3})$ 
when $\langle\phi\, ,\, \psi_{{}_{0}}
\rangle_{\mathcal H} \ne 0$, 
on the other hand, 
the right hand side of (\ref{eq:identity}) 
is in $L^{2}({\mathbb R}^{3})$. 
Let us consider the case where 
$\langle\phi\, ,\, \psi_{{}_{0}}
\rangle_{\mathcal H} = 0$ now.  
In this case, since we took an arbitrary $\phi$ from 
$D(I\otimes N_{\mathrm f}^{1/2})$ which is dense 
in $L^{2}({\mathbb R}^{3})$, 
we have ${\psi_{{}_{0}}} = 0$, which also implies 
a contradiction. 
Therefore, we obtain Theorem \ref{theorem:2}.

\section{Sharp estimate of total number of soft bosons}
\label{section:EESPB}

In this section, we prove Theorem \ref{theorem:1}. 
So, we assume $\kappa > 0$ throughout this section. 
In order to prove Theorem \ref{theorem:1}, we justify 
the following symbolic identity  
\begin{eqnarray}
\langle \psi_{\kappa}\, ,\, 
I\otimes N_{\mathrm f}\psi_{\kappa}
\rangle_{\mathcal H} = 
\int_{{\mathbb R}^{3}}\| 
I\otimes a(k)\psi_{\kappa}\|_{\mathcal H}^{2}
d^{3}k. 
\label{eq:photon-numbers}
\end{eqnarray}

Let $X = (X , {\mathcal A}, \mu)$ be a $\sigma$-finite 
measurable space. Define the symmetric 
Fock space ${\mathcal F}_{X}$ 
from $X$ by  
$${\mathcal F}_{X} = \bigoplus_{n=0}^{\infty}
\otimes^{n}_{s}L^{2}(X).$$ 
The annihilation operator $a(f)$, $f \in L^{2}(X)$, 
and the number operator $N$ acting in ${\mathcal F}_{X}$  
can be defined in the same way as in 
(\ref{eq:smeard-annihilation}) and 
(\ref{eq:number-operator-rigorous}) for those acting in 
${\mathcal F}$, respectively.

\begin{proposition}
\label{prp:advised-by-referee}
For arbitrary complete orthonormal system 
$\left\{ f_{\nu}\right\}_{\nu}$ of $L^{2}(X)$, 
\begin{eqnarray}
\| N^{1/2}\Psi\|_{{\mathcal F}_{X}}^{2} 
= \sum_{\nu=1}^{\infty}\| a(f_{\nu})\Psi
\|_{{\mathcal F}_{X}}^{2}, \qquad 
\forall \Psi \in  D(N^{1/2}).  
\label{eq:advised-by-referee}
\end{eqnarray}
\end{proposition}

\proof 
Set 
\begin{eqnarray*}
&{}& 
\Psi^{(n)}_{M}(k_{1},\cdots,k_{n}) 
= \sum_{\nu=1}^{M}
\biggl| \left( f_{\nu}\, ,\, \Psi^{(n+1)}
(\cdot,k_{1},\cdots,k_{n})
\right)_{L^{2}(X)}\biggr|^{2}, \\ 
&{}& 
d\mu^{n}(k_{1},\cdots,k_{n}) = d\mu(k_{1})\otimes 
\cdots\otimes d\mu(k_{n}). 
\end{eqnarray*}
By the definition of the annihilation operator, 
for each $M \in  {\mathbb N}$ and every $\Psi 
\in  D(N^{1/2})$ we have 
$$\sum_{\nu=1}^{M}\| a(f_{\nu})\Psi\|_{{\mathcal F}_{X}}^{2} 
= \sum_{n=0}^{\infty}(n+1)\int_{X^{n}}\Psi^{(n)}_{M}
(k_{1},\cdots,k_{n})d\mu^{n}(k_{1},\cdots,k_{n}).$$
Since $\Psi^{(n+1)}(\cdot, k_{1},\cdots, k_{n}) 
\in  L^{2}(X)$ for $\mu^{n}$-a.e. $(k_{1},\cdots, k_{n}) 
\in X^{n}$, we have 
\begin{eqnarray*}
\Psi^{(n)}_{M}(k_{1},\cdots, k_{n}) 
\le 
\| \Psi^{(n+1)}(\cdot, k_{1},\cdots, k_{n})\|_{
L^{2}(X)}^{2}, \,\,\, 
\mu^{n}\mbox{-a.e.}\, (k_{1},\cdots, k_{n}) 
\in X^{n}, 
\end{eqnarray*}
by Bessel's inequality. 
Since $\left\{ f_{\nu}\right\}_{\nu}$ is complete, 
$\Psi^{(n)}_{M}(k_{1},\cdots, k_{n})$ converges to  \\ 
$\|\Psi^{(n+1)}(\cdot, k_{1},\cdots, k_{n})\|_{
L^{2}(X)}^{2}$ as $M\to\infty$. 
Therefore, (\ref{eq:advised-by-referee}) 
follows from Lebesgue's 
monotone convergence theorem.  
\qed

\begin{lemma}
\label{lmm:9/9-1}
For every $\kappa$ with $ 0 < \kappa < \Lambda$, 
\begin{eqnarray*}
\psi_{\kappa} \in  D(I\otimes N_{\mathrm f}). 
\end{eqnarray*}
\end{lemma}

\proof
Let 
${\mathbb R}^{3}_{\le\kappa} = 
\left\{ k \in  {\mathbb R}^{3}\, |\, |k| \le \kappa\right\}$ 
and 
${\mathbb R}^{3}_{>\kappa} = 
\left\{ k \in  {\mathbb R}^{3}\, |\, |k| > \kappa\right\}$. 
We set $N_{\mathrm f}^{\le\kappa} = d\Gamma( 
{1\!\!\!\>\!\>{\rm l}}_{\left[ 0 , \kappa\right]})$ 
and $N_{\mathrm f}^{>\kappa} = d\Gamma( 
{1\!\!\!\>\!\>{\rm l}}_{\left( \kappa,\infty\right)})$ 
acting in $\bigoplus_{n=0}^{\infty}\otimes_{\mathrm s}^{n}
L^{2}({\mathbb R}^{3}_{\le\kappa})$ and 
$\bigoplus_{n=0}^{\infty}\otimes_{\mathrm s}^{n}
L^{2}({\mathbb R}^{3}_{>\kappa})$, respectively. 
We note 
\begin{eqnarray}
D((H_{\mathrm f}^{>\kappa})^{s}) 
\subset D((N_{\mathrm f}^{>\kappa})^{s})
\label{eq:free<number}
\end{eqnarray}
for $s > 0$. 
Through the unitary equivalence 
$H^{\mbox{\tiny N}}_{\kappa} \cong 
H_{\mathrm f}^{\le \kappa}\otimes I + 
I \otimes H_{\mbox{\tiny N}}^{>\kappa}$, 
the ground state $\psi_{\kappa}$ of 
$H^{\mbox{\tiny N}}_{\kappa}$ is represented 
by $\Omega_{0}\otimes\psi_{\kappa}^{>\kappa}$, 
where $\psi_{\kappa}^{>\kappa}$ is a 
ground state of $H_{\mbox{\tiny N}}^{>\kappa}$ 
and $\Omega_{0}$ the Fock vacuum. 
We note $N_{\mathrm f} \cong N_{\mathrm f}^{\le\kappa}\otimes I 
+ I\otimes N_{\mathrm f}^{>\kappa}$. 
Since $D(H_{\mathrm f}^{> \kappa}) \subset  
D(N_{\mathrm f}^{> \kappa})$ by (\ref{eq:free<number}), 
$\psi_{\kappa}^{>\kappa}$ is in 
$D(N_{\mathrm f}^{>\kappa})$, i.e.,  
$\psi_{\kappa}^{>\kappa} \in  D(N_{\mathrm f}^{>\kappa})$, 
by Proposition \ref{proposition:nelson-sa}.  
Hence our lemma follows. 
\qed

Setting $X = {\mathbb R}^{3}$ in Proposition 
\ref{prp:advised-by-referee} and using the 
identification (\ref{eq:identification}) and 
Lemma \ref{lmm:9/9-1}, we obtain mathematical 
justification of (\ref{eq:photon-numbers}): 

\begin{corollary}
\label{cor:photon-number-rigorous-kore}
For every $\kappa$ with $0 < \kappa < \Lambda$ 
and an arbitrary complete orthonormal system 
$\left\{ f_{\nu}\right\}_{\nu = 1}^{\infty}$ 
of $L^{2}({\mathbb R}^{3})$, 
\begin{eqnarray} 
\langle\psi_{\kappa}\, ,\, 
I\otimes 
N_{\mathrm f}\psi_{\kappa}\rangle_{\mathcal H} 
= \| I\otimes 
N_{\mathrm f}^{1/2}\psi_{\kappa}\|_{\mathcal H}^{2}  
= \sum_{\nu=1}^{\infty}
\| I\otimes 
a(f_{\nu})\psi_{\kappa}\|_{\mathcal H}^{2}.
\label{eq:photon-number-rigorous-kore}
\end{eqnarray}
\end{corollary}

\qquad 

\underbar{\it Proof of Theorem \ref{theorem:1}}: 
Fix $\kappa$ satisfying $0 < \kappa < \Lambda$. 
We assume all hypotheses of Theorem \ref{theorem:1}. 
Let $\left\{ f_{\nu}\right\}_{\nu = 1}^{\infty} 
\subset C_{0}^{\infty}({\mathbb R}^{3}
\setminus \left\{ 0\right\})$ 
and $\left\{ e_{p}\right\}_{p=1}^{\infty}$ 
be complete orthonormal systems of $L^{2}({\mathbb R}^{3})$ 
and ${\mathcal H}$, respectively. 
Then, $\left\{ f_{\nu}(\cdot)e_{p}\right\}_{\nu,p=1}^{\infty}$ 
is a complete orthonormal system of 
$L^{2}({\mathbb R}^{3} ; {\mathcal H})$. 
By using Parseval's equality, 
we have
\begin{eqnarray}
\int_{{\mathbb R}^{3}}\| J_{j}(k)\psi_{\kappa}
\|_{\mathcal H}^{2}d^{3}k 
&\equiv& 
\| J_{j}(\cdot)\psi_{\kappa}\|_{L^{2}({\mathbb R}^{3} ; 
{\mathcal H})}^{2} 
\label{eq:12-16-1} \\ 
\nonumber 
&=& 
\sum_{\nu=1}^{\infty}
\sum_{p=1}^{\infty}
\Biggl| 
\int_{{\mathbb R}^{3}}
\langle f_{\nu}(k)e_{p}\, ,\, 
J_{j}(k)\psi_{\kappa}
\rangle_{\mathcal H}d^{3}k
\Biggr|^{2} \\ 
\nonumber 
&=& 
\sum_{\nu=1}^{\infty}
\sum_{p=1}^{\infty}
\Biggl| 
\langle e_{p}\, ,\, 
\int_{{\mathbb R}^{3}}
f_{\nu}(k)^{*}J_{j}(k)\psi_{\kappa}d^{3}k
\rangle_{\mathcal H}
\Biggr|^{2} \\ 
\nonumber 
&=& 
\sum_{\nu=1}
\Biggl\|
\int_{{\mathbb R}^{3}}
f_{\nu}(k)^{*}
J_{j}(k)\psi_{\kappa}d^{3}k
\Biggr\|_{\mathcal H}^{2}
\end{eqnarray}
for $j = 1, 2$ 
since $J_{j}(\cdot)\psi_{\kappa} 
\in  L^{2}({\mathbb R}^{3} ; {\mathcal H})$ 
for every $\kappa$ satisfying $0 < \kappa < \Lambda$. 
Applying the triangle inequality to 
(\ref{eq:a-representation-simple}) and using 
(\ref{eq:photon-number-rigorous-kore}) 
and (\ref{eq:12-16-1}), we have  
\begin{eqnarray}
&{}& 
\langle \psi_{\kappa}\, ,\, 
I\otimes N_{\mathrm f}\psi_{\kappa}
\rangle_{\mathcal H} = 
\sum_{\nu=1}^{\infty} 
\|I\otimes a(f_{\nu})\psi_{\kappa}\|_{\mathcal H}^{2}
\label{eq:T1-(1)}  \\ 
\nonumber 
&{}&\le 
2 
\int_{{\mathbb R}^{3}}\| J_{1}(k)\psi_{\kappa}\|_{\mathcal H}^{2} 
+ 
2
\int_{{\mathbb R}^{3}}\| J_{2}(k)\psi_{\kappa}\|_{\mathcal H}^{2}. 
\end{eqnarray}
By (\ref{eq:estimate-J1}), 
(\ref{eq:estimate-J2}), (\ref{eq:T1-(1)}), we have 
\begin{eqnarray} 
\qquad 
\langle \psi_{\kappa}\, ,\, 
I\otimes N_{\mathrm f}\psi_{\kappa}
\rangle_{\mathcal H} 
\le 
2\left\{ 
\frac{\textsl{q}^{2}}{4\pi^{2}}\left( 
\log\frac{\Lambda}{\kappa}\right) + 
\frac{\textsl{q}^{2}}{8\pi^{2}}\Lambda^{2}
\| |x|\otimes I\psi_{\kappa}\|_{\mathcal H}^{2}
\right\}. 
\label{eq:T1-(3)}
\end{eqnarray}
By (\ref{eq:a-representation-simple}) again, 
we have 
$$\int_{{\mathbb R}^{3}}
f_{\nu}(k)^{*}J_{1}(k)\psi_{\kappa}d^{3}k 
= I\otimes a(f_{\nu})\psi_{\kappa} 
- \int_{{\mathbb R}^{3}}
f_{\nu}(k)^{*}J_{2}(k)\psi_{\kappa}d^{3}k.$$ 
In the same way as above, we get 
\begin{eqnarray} 
\int_{{\mathbb R}^{3}}
\| J_{1}(k)\psi_{\kappa}\|_{\mathcal H}^{2}d^{3}k
&\le&  
2\langle\psi_{\kappa}\, ,\, 
I\otimes N_{\mathrm f}\psi_{\kappa}
\rangle_{\mathcal H} 
+ 
2 
\int_{{\mathbb R}^{3}}
\| J_{2}(k)\psi_{\kappa}\|_{\mathcal H}^{2}d^{3}k.
\label{eq:T1-(4)}    
\end{eqnarray}
By (\ref{eq:estimate-J1}), (\ref{eq:estimate-J2}), 
and (\ref{eq:T1-(4)}),  
we have 
$$\frac{\textsl{q}^{2}}{4\pi^{2}}
\left(\log\frac{\Lambda}{\kappa}\right)
\le  2\langle \psi_{\kappa}\, ,\, 
I\otimes N_{\mathrm f}\psi_{\kappa}
\rangle_{\mathcal H} 
+ 2
\frac{\textsl{q}^{2}}{8\pi^{2}}\Lambda^{2}
\| |x|\otimes I\psi_{\kappa}\|_{\mathcal H}^{2},
$$
which implies 
\begin{eqnarray}
\qquad 
\left\{ 
\frac{\textsl{q}^{2}}{8\pi^{2}}
\left(\log\frac{\Lambda}{\kappa}\right) 
- \frac{\textsl{q}^{2}}{8\pi^{2}}\Lambda^{2}
\| |x|\otimes I\psi_{\kappa}\|_{\mathcal H}^{2}
\right\} 
\le 
\langle \psi_{\kappa}\, ,\, 
I\otimes N_{\mathrm f}\psi_{\kappa}
\rangle_{\mathcal H}.
\label{eq:T1-(5)}
\end{eqnarray}
Therefore, 
(\ref{eq:ine1}) follows from 
(\ref{eq:T1-(3)}) and (\ref{eq:T1-(5)}).

\section{Finite uncertainty of position in 
ground state} 
\label{section:FNPGS}

In this section, we show that if 
$H_{\kappa}^{\mbox{\tiny N}}$ has a (normalized) 
ground state, then the uncertainty of the position 
in the ground state has to be finite. 
More precisely, if $H_{\kappa}^{\mbox{\tiny N}}$ 
has a ground state $\psi_{\kappa}$, 
then $\psi_{\kappa} \in  D(x^{2}\otimes I)$. 
Therefore, contrary to (\ref{eq:uncertainty-position}), 
we can indirectly prove that uncertainty 
of the position in the ground state is finite, 
$(\Delta x)_{\mathrm {gs}} < \infty$.

In the first half of this section, we consider 
the case where $V$ is in class (C1) and prove 
that if $H_{\kappa}^{\mbox {\tiny N}}$ has a 
ground state $\psi_{\kappa}$, 
then $\psi_{\kappa}$ belongs 
to $D(x^{2}\otimes I)$. 
Moreover, to prove Theorem \ref{crl:1},  
we need a uniform estimate of 
$\| |x|\otimes I\psi_{\kappa}\|_{\mathcal H}$ 
in the infrared cutoff $\kappa$. 
To do that we prepare some inequalities.

\begin{lemma}
\label{lemma:norm-estimate0}
Assume (A). Then, there exists a constant 
$C_{\textsl{q}} > 0$ such that
\begin{eqnarray}
\sup_{0 < \kappa < \Lambda}
\|\left( H_{\mathrm 0} + I\right)
\left( H^{\mbox{\rm {\tiny N}}}_{\kappa} 
- E^{\mbox{\rm {\tiny N}}}_{\kappa} + I\right)^{-1}\| 
\le C_{\textsl{q}}. 
\label{eq:norm-estimate0-k}
\end{eqnarray}
\end{lemma}

\proof 
For every $L^{2}({\mathbb R}^{3})$-valued function 
$f_{x} : {\mathbb R}^{3}_{x} \to L^{2}({\mathbb R}^{3})$ 
(i.e., $f_{x} \in  L^{2}({\mathbb R}^{3})$ for a.e. 
$x \in  {\mathbb R}^{3}$ and $\|f_{\star}\|_{L^{2}} 
\in  L^{2}({\mathbb R}^{3})$), we set 
$\| f_{\star}\|_{L^{2},\infty}$ $:=$ 
${\mathrm {ess}}.\sup_{x\in{\mathbb R}^{3}}$
$\| f_{x}\|_{L^{2}}$. 
Combining fundamental inequalities for $H_{\mathrm {f}}$ 
and $\phi_{\kappa}(x)$ with an argument on the constant 
fiber direct integral (see, e.g., \cite[Lemma 13-12]{arai-text}), 
for every $\varepsilon, 
\varepsilon' > 0$ and every $\psi \in  D(H_{0})$, 
we have 
\begin{eqnarray}
\| H_{{\mathrm {I},\kappa}}\psi\|_{\mathcal H}^{2} 
&\le& 
(2+\varepsilon)
\|\sqrt{2}\lambda_{\kappa,\star}/\sqrt{\omega}
\|_{L^{2},\infty}^{2}
\| I\otimes H_{\mathrm {f}}^{1/2}\psi\|_{\mathcal H}^{2} 
\label{eq:4-18-1}  \\ 
\nonumber 
&{}& \qquad\qquad\qquad 
+ 
\frac{1}{2}\left( 1 + \frac{1}{2\varepsilon}\right)
\|\sqrt{2}\lambda_{\kappa,\star}
\|_{L^{2},\infty}^{2}
\|\psi\|_{\mathcal H}^{2}  \\ 
\nonumber 
&=& 
(2+\varepsilon)
\|\sqrt{2}\lambda_{\kappa,0}/\sqrt{\omega}
\|_{L^{2}}^{2}
\| I\otimes H_{\mathrm {f}}^{1/2}\psi\|_{\mathcal H}^{2} \\ 
\nonumber 
&{}& \qquad\qquad\qquad
+ 
\frac{1}{2}\left( 1 + \frac{1}{2\varepsilon}\right)
\|\sqrt{2}\lambda_{\kappa,0}
\|_{L^{2}}^{2}
\|\psi\|_{\mathcal H}^{2}, 
\end{eqnarray}
since $|e^{-ikx}| = 1$. 
By fundamental inequalities, we have 
\begin{eqnarray}
\| I\otimes H_{\mathrm {f}}^{1/2}\psi\|_{\mathcal H}^{2} 
&=& 
\langle \psi\, ,\, I\otimes H_{\mathrm {f}}\psi
\rangle_{\mathcal H} 
\le \|\psi\|_{\mathcal H}
\| I\otimes H_{\mathrm {f}}\psi\|_{\mathcal H} 
\label{eq:4-18-2}  \\ 
\nonumber 
&\le& 
\varepsilon' 
\| (H_{0} + I)\psi\|_{\mathcal H}^{2} 
+ \frac{1}{4\varepsilon'}
\|\psi\|_{\mathcal H}^{2}.  
\end{eqnarray}
It follows from direct estimates that 
\begin{eqnarray}
\| \lambda_{\kappa,0}\|_{L^{2}}^{2} 
\le \frac{\Lambda^{2}}{8\pi^{2}}, 
\qquad 
\| \lambda_{\kappa,0}/\sqrt{\omega}\|_{L^{2}}^{2} 
\le \frac{\Lambda}{4\pi^{2}}. 
\label{eq:4-18-3}
\end{eqnarray}
By (\ref{eq:4-18-1}) -- (\ref{eq:4-18-3}), 
we have 
\begin{eqnarray}
\| H_{{\mathrm I},\kappa}\psi\|_{\mathcal H} 
\le 
C_{\Lambda}^{(1)}(\varepsilon,\varepsilon')
\| (H_{0} + I)\psi\|_{\mathcal H} 
+ 
C_{\Lambda}^{(2)}(\varepsilon,\varepsilon')
\| \psi\|_{\mathcal H}, 
\label{eq:4-18-4}
\end{eqnarray}
where 
\begin{eqnarray*}
&{}& 
C_{\Lambda}^{(1)}(\varepsilon,\varepsilon') 
= \frac{\sqrt{\Lambda}}{2\pi}
\sqrt{2\varepsilon'(2+\varepsilon)}, \\ 
&{}& 
C_{\Lambda}^{(2)}(\varepsilon,\varepsilon') 
= \frac{\sqrt{\Lambda}}{2\pi}
\sqrt{\frac{2+\varepsilon}{2\varepsilon'} 
+ \frac{1}{2}\left(1 + \frac{1}{2\varepsilon}
\right)\Lambda}. 
\end{eqnarray*}
Since $(H_{0} + I)\psi 
= (H_{\kappa}^{\mbox{\tiny N}} 
- E_{\kappa}^{\mbox{\tiny N}} + I)\psi 
- \textsl{q}H_{{\mathrm I},\kappa}\psi 
+ (E_{\kappa}^{\mbox{\tiny N}} - E_{\mathrm {at}})\psi$ 
for every $\psi \in  D(H_{0})$ and 
$|E_{\kappa}^{\mbox{\tiny N}} - E_{\mathrm {at}}| 
\le \textsl{q}^{2}\|\lambda_{\kappa,0}\|^{2}_{L^{2}}$ by 
Proposition \ref{proposition:nelson-sa}, 
\begin{eqnarray*}
\| (H_{0} + I)\psi\|_{\mathcal H} 
&\le&  
\frac{1}{1 - 
|\textsl{q}|C_{\Lambda}^{(1)}(\varepsilon,\varepsilon')}
\| (H_{\kappa}^{\mbox{\tiny N}} 
- E_{\kappa}^{\mbox{\tiny N}} + I)\psi\|_{\mathcal H} \\ 
&{}& \qquad 
+ 
\frac{|\textsl{q}|C_{\Lambda}^{(2)}(\varepsilon,\varepsilon') 
+ \textsl{q}^{2}\Lambda^{2}/8\pi^{2}}{1 - 
|\textsl{q}|C_{\Lambda}^{(1)}(\varepsilon,\varepsilon')}
\|\psi\|_{\mathcal H}
\end{eqnarray*}
for every $\varepsilon, \varepsilon' > 0$ 
satisfying $1 - 
|\textsl{q}|C_{\Lambda}^{(1)}(\varepsilon,\varepsilon') > 0$, 
which implies 
\begin{eqnarray*}
\| (H_{0} + I)
(H_{\kappa}^{\mbox{\tiny N}} 
- E_{\kappa}^{\mbox{\tiny N}} + I)^{-1}\| 
&\le&  
\frac{1 + 
|\textsl{q}|C_{\Lambda}^{(2)}(\varepsilon,\varepsilon') 
+ \textsl{q}^{2}\Lambda^{2}/8\pi^{2}}{1 - 
|\textsl{q}|C_{\Lambda}^{(1)}(\varepsilon,\varepsilon')}.
\end{eqnarray*}
\qed

We obtain the following lemma from 
Lemma \ref{lemma:norm-estimate0}.

\begin{lemma}
\label{lemma:norm-estimate} 
For every $\textsl{q} \ne 0$ 
and arbitrary $\kappa, \epsilon$ with 
$0 < \epsilon$ and $0 \le \kappa < \Lambda$,   
\begin{eqnarray} 
\|( H_{\mathrm 0} + I)
( H^{\mbox{\rm {\tiny N}}}_{\kappa} 
- E^{\mbox{\rm {\tiny N}}}_{\kappa} + \epsilon)^{-1}\| 
\le 
\frac{\displaystyle 
C_{\textsl{q}}}{\displaystyle  
\min\left\{\epsilon,\,\,\, 1\right\}}.  
\label{eq:norm-estimate}
\end{eqnarray}
\end{lemma}

\proof 
(\ref{eq:norm-estimate}) follows from 
(\ref{eq:norm-estimate0-k}) 
and 
$$
\|( H^{\mbox{\tiny N}}_{\kappa} 
- E^{\mbox{\tiny N}}_{\kappa} + I)
( H^{\mbox{\tiny N}}_{\kappa} 
- E^{\mbox{\tiny N}}_{\kappa} + \epsilon)^{-1}\| 
\le 
\left\{ 
  \begin{array}{@{\,}ll}
\epsilon^{-1}
& \mbox{if $\epsilon < 1$,} \\  
\qquad & \mbox{\qquad} \\ 
1
& \mbox{if $\epsilon \ge 1$.} 
\end{array}
\right.   
$$
\qed

\qquad

\begin{lemma}
\label{lemma:|x|-estimate1} 
Let $V$ be in class (C1). 
\begin{description}
\item[(i)] $D(V) \subset D(x^{2})$. 
\item[(ii)] 
If $\psi$ is in $D( H_{\mathrm 0})$, 
then $\psi \in D(|x|\otimes I)$ and 
\end{description}
\begin{eqnarray} 
\qquad 
\||x|\otimes I \psi\|_{\mathcal H}^{2} 
\le c_{1}\|H_{\mathrm 0}^{1/2}\psi\|_{\mathcal H}^{2} 
+ c_{2}\|\psi\|_{\mathcal H}^{2} 
\le 
c_{1}\|H_{\mathrm 0}\psi\|_{\mathcal H}^{2} 
+ (c_{1} + c_{2})\|\psi\|_{\mathcal H}^{2}.
\label{eq:|x|-estimate1-1}
\end{eqnarray}
\begin{description}
\item[\hspace{5mm}]
In particular, for $0 \le \kappa \le \Lambda$
\end{description}
\begin{eqnarray}
\qquad 
\||x|\otimes I \psi_{\kappa}\|_{\mathcal H}^{2} 
\le   
\left( 
c_{1}C_{\textsl{q}}^{2} + c_{1} + 
c_{2}\right).  
\label{eq:|x|-estimate1-2}
\end{eqnarray}
\end{lemma}

\proof 
(i) directly follows from the first inequality of (C1-2).  
We obtain the first statement of (ii) by (C1-1), 
Proposition \ref{proposition:nelson-sa}, and (i). 
As for the second statement, 
the first inequality of (\ref{eq:|x|-estimate1-1}) 
is obtained in the same way as in \cite[Lemma 4.6]{arai}. 
By Schwarz' inequality, we have 
\begin{eqnarray*}
&{}& \| H_{\mathrm 0}^{1/2}\psi\|_{\mathcal H}^{2} 
= \langle\psi\, ,\, H_{\mathrm 0}\psi\rangle_{\mathcal H} 
\le \|\psi\|_{\mathcal H}\| H_{\mathrm 0}\psi\|_{\mathcal H} 
\le \| H_{\mathrm 0}\psi\|_{\mathcal H}^{2} 
+ \frac{1}{4}\|\psi\|_{\mathcal H}^{2} \\ 
&{}& 
\le \| H_{\mathrm 0}\psi\|_{\mathcal H}^{2} 
+ \|\psi\|_{\mathcal H}^{2} 
\end{eqnarray*} 
for $\psi \in  D(H_{\mathrm 0})$. 
So, we obtain the second inequality of (\ref{eq:|x|-estimate1-1}). 
By (\ref{eq:|x|-estimate1-1}), 
we have 
\begin{eqnarray*} 
\||x|\otimes I \psi_{\kappa}\|_{\mathcal H}^{2} 
&\le& 
c_{1}\|( H_{\mathrm 0} + I)( H^{\mbox{\tiny N}}_{\kappa} 
- E^{\mbox{\tiny N}}_{\kappa} + I)^{-1}
( H^{\mbox{\tiny N}}_{\kappa} 
- E^{\mbox{\tiny N}}_{\kappa} + I)\psi_{\kappa}\|_{\mathcal H}^{2} \\ 
&{}&   
+ (c_{1} + c_{2}) \\ 
&=& 
c_{1}\|( H_{\mathrm 0} + I)( H^{\mbox{\tiny N}}_{\kappa} 
- E^{\mbox{\tiny N}}_{\kappa} + I)^{-1}
\psi_{\kappa}\|_{\mathcal H}^{2} 
+ (c_{1} + c_{2}). 
\end{eqnarray*}
This inequality and Lemma \ref{lemma:norm-estimate} 
imply (\ref{eq:|x|-estimate1-2}). 
\qed

The following proposition 
follows from Lemma \ref{lemma:|x|-estimate1}  
directly: 

\begin{proposition}
[finite uncertainty of position in ground state]
\label{prp:|x|^{2}-domain-1}
Let $V$ be in class (C1) 
and $\kappa$ satisfy $0 \le \kappa < \Lambda$. 
If $H_{\kappa}^{\mbox{\rm {\tiny N}}}$ has 
a ground state $\psi_{\kappa}$, then 
$\psi_{\kappa} \in  D(x^{2}\otimes I)$. 
Moreover, $\sup_{0<\kappa<\Lambda}
\| |x|\otimes I \psi_{\kappa}\|_{\mathcal H} 
< \infty$, provided that $\psi_{\kappa}$ 
exists for $0 < \kappa < \Lambda$. 
\end{proposition}

\proof
Suppose that there exists a ground state 
$\psi_{\kappa}$ of $H_{\kappa}^{\mbox{\rm {\tiny N}}}$. 
Then, by (C1-1), 
Lemma \ref{lemma:|x|-estimate1}(i), and 
Proposition \ref{proposition:nelson-sa}, 
we have $\psi_{\kappa} \in D(H_{\kappa}^{\mbox{\rm {\tiny N}}}) 
\subset D(H_{\mathrm {at}}\otimes I) 
\subset D(x^{2}\otimes I)$.  
The uniform estimate of $\| |x|\otimes I 
\psi_{\kappa}\|_{\mathcal H}$ in $\kappa$ follows 
from (\ref{eq:|x|-estimate1-2}) directly. 
\qed

In the last half of this section, we consider 
the case where $V$ is in class (C2) and 
we prove that if $H_{\kappa}^{\mbox {\tiny N}}$ has a 
ground state $\psi_{\kappa}$, 
then $\psi_{\kappa}$ belongs 
to $D(x^{2}\otimes I)$. 
Moreover, we show a uniform estimate of 
$\| |x|\otimes I\psi_{\kappa}\|_{\mathcal H}$ 
in the infrared cutoff $\kappa$, 
by proving the so-called exponential decay.

Let ${E^{\mbox{\tiny N}}_{\kappa}}^{V=0} = \inf\sigma 
\left( {H^{\mbox{\tiny N}}_{\kappa}}^{V=0}\right)$, 
where the superscript of ${H^{\mbox{\tiny N}}_{\kappa}}^{V=0}$ 
means that in (\ref{eq:Nelson-Hamiltonian}) 
the external potential $V$ is omitted. 
The (positive) binding energy is defined by 
\begin{eqnarray*}
E_{\kappa}^{\mathrm {bin}} 
:= {E^{\mbox{\tiny N}}_{\kappa}}^{V=0} 
- E^{\mbox{\tiny N}}_{\kappa}. 
\label{def-binding-energy} 
\end{eqnarray*} 

The binding energy is bounded from below:  

\begin{proposition}[strict positivity of binding energy]
\label{proposition:binding-energy} 
Let $V$ be in class (C2).  
Fix $\kappa$ with $0 \le \kappa < \Lambda$. 
Then,  
\begin{eqnarray}
E_{\kappa}^{\mathrm {bin}} \ge - {E_{\mathrm {at}}} > 0.
\label{eq:8-4-1}
\end{eqnarray}
\end{proposition}

\proof 
Using the idea proving \cite[Theorem 3.1]{gll} 
for the Pauli-Fierz model, 
relation (\ref{eq:8-4-1}) was proved in 
\cite[Proposition 4.4]{hhs}, but for the special 
external potential.  
It is easy to see that 
our proposition is also proven in the same way as 
in \cite[Proposition 4.4]{hhs} 
following the idea in the proof of \cite[Theorem 3.1]{gll}. 
\qed

\begin{proposition}[exponential decay]
\label{new-exp-decay}  
Fix $\kappa$ with $0 \le \kappa < \Lambda$. 
Let $V$ be in class (C2). 
Assume $H^{\mbox{\rm {\tiny N}}}_{\kappa}$ has a 
(normalized) ground state $\psi_{\kappa}$.  
Then, there exist a sufficiently small 
$C_{0} > 0$, a sufficiently large $N_{0} \in  {\mathbb N}$, 
and $C > 0$ 
such that $\psi_{\kappa} 
\in  D(e^{C_{0}|x|})$ and 
\begin{eqnarray}
&{}& \| e^{C_{0}|x|}\psi_{\kappa}\|_{\mathcal H} 
\label{eq:new-1} \\ 
\nonumber 
&\le& 
e^{3C_{0}N_{0}}\left\{ 
1 + C 
\left(
|E_{\mathrm {at}}| - \sup_{N_{0}<|x|}|V(x)| 
- C_{0}^{2}
\right)^{-1/2}
\right\},  
\end{eqnarray}
where 
\begin{eqnarray}
|E_{\mathrm {at}}| - \sup_{N_{0}<|x|}|V(x)| 
- C_{0}^{2} > 0. 
\label{eq:new-2}
\end{eqnarray}
\end{proposition}

\proof 
Since $\lim_{|x|\to\infty}|V(x)| = 0$ in (C2-1), 
we can take $N_{0} \in  {\mathbb N}$ and 
$C_{0} > 0$ such that 
(\ref{eq:new-2}) holds because 
we assumed $E_{\mathrm {at}} < 0$ in (C2-2). 
We take a non-negative function 
${1\!\!\!\>\!\>{\rm l}}_{n} \in  C_{0}^{\infty}({\mathbb R})$ 
for each $n \in {\mathbb N}$ 
satisfying ${1\!\!\!\>\!\>{\rm l}}_{n}(r) = 1$ for $|r| \le n$;\, 
$=0$ for $|r| \ge 3n$, 
$0 \le {1\!\!\!\>\!\>{\rm l}}_{n}(r) \le 1$ for 
$n < |r| < 3n$. 
Since ${1\!\!\!\>\!\>{\rm l}}_{n}' \in  C_{0}^{\infty}({\mathbb R})$ 
again, we have  
$C_{n} := \sup_{r}
|d{1\!\!\!\>\!\>{\rm l}}_{n}(r)/dr| < \infty$. 
We set $f_{\varepsilon}(r) := r(1+\varepsilon r)^{-1}$ 
for every $\varepsilon > 0$ and $r \ge 0$. 
We define a function $G_{n,\varepsilon}(x)$ by 
$G_{n,\varepsilon}(x) := (1 - {1\!\!\!\>\!\>{\rm l}}_{n}(|x|))
f_{\varepsilon}(e^{C_{0}|x|})$. 
Since $0 \le f_{\varepsilon}(r) \le \varepsilon^{-1}$ 
for all $r \ge 0$, 
the multiplication operators 
$f_{\varepsilon}(e^{C_{0}|x|})$ 
and $G_{n,\varepsilon}$ are bounded on 
$L^{2}({\mathbb R}^{3})$.    
In the same way as in \cite[Lemma 5.1]{hhs}, 
we have 
\begin{eqnarray}
E_{\kappa}^{\mathrm {bin}} 
\| G_{n,\varepsilon}\otimes I\psi_{\kappa}\|_{\mathcal H}^{2} 
&\le& \frac{1}{2}
\langle \psi_{\kappa}\, ,\, 
|\nabla G_{n,\varepsilon}|^{2}\otimes I \psi_{\kappa}
\rangle_{\mathcal H}
\label{eq:exp-decay-1} \\ 
\nonumber 
&{}& 
+ \sup_{n<|x|}|V(x)|
\langle \psi_{\kappa}\, ,\, 
G_{n,\varepsilon}^{2}\otimes I \psi_{\kappa}
\rangle_{\mathcal H}.
\end{eqnarray}
It is easy to check that 
\begin{eqnarray*}
\frac{\partial G_{n,\varepsilon}(x)}{\partial x_{j}} 
&=&  
- \frac{\partial {1\!\!\!\>\!\>{\rm l}}_{n}
(|x|)}{\partial x_{j}}
f_{\varepsilon}(e^{C_{0}|x|}) 
+ 
C_{0}(1 - {1\!\!\!\>\!\>{\rm l}}_{n}(|x|))
\frac{e^{C_{0}|x|}}{
\left( 1 + \varepsilon e^{C_{0}|x|}\right)^{2}} 
\frac{x_{j}}{|x|}.
\end{eqnarray*}
So, using ${\mathrm {supp}}
{1\!\!\!\>\!\>{\rm l}}_{n}' \subset 
\left[ - 3n , -n\right]\cup 
\left[ n , 3n\right]$ and 
$( 1 + \varepsilon e^{C_{0}|x|})^{-4} 
< ( 1 + \varepsilon e^{C_{0}|x|})^{-2}$, 
we have 
\begin{eqnarray}
|\nabla G_{n,\varepsilon}(x)|^{2} 
&\le& 
2 \left(
\sup_{n\le |x|\le 3n}f_{\varepsilon}
(e^{C_{0}|x|})\right)^{2}
\sum_{j=1}^{3}\left(
\frac{\partial 
{1\!\!\!\>\!\>{\rm l}}_{n}(|x|)}{\partial x_{j}}\right)^{2} 
\label{eq:exp-decay-2} \\ 
\nonumber 
&{}& 
+ 2C_{0}^{2}
(1 - {1\!\!\!\>\!\>{\rm l}}_{n}(|x|))^{2}
\frac{e^{2C_{0}|x|}}{\left( 
1 + \varepsilon e^{C_{0}|x|}
\right)^{4}} \\
\nonumber 
&\le& 
2\left(
\frac{e^{3C_{0}n}}{1 
+ \varepsilon e^{3C_{0}n}}
\right)^{2} 
\sum_{j=1}^{3}\left(
\frac{\partial 
{1\!\!\!\>\!\>{\rm l}}_{n}(|x|)}{\partial x_{j}}\right)^{2} 
+ 2C_{0}^{2}G_{n,\varepsilon}(x)^{2} \\ 
\nonumber 
&\le& 
2e^{6C_{0}n}
\sum_{j=1}^{3}\left(
\frac{\partial 
{1\!\!\!\>\!\>{\rm l}}_{n}(|x|)}{\partial x_{j}}\right)^{2} 
+ 2C_{0}^{2}G_{n,\varepsilon}(x)^{2}.
\end{eqnarray}
It is easy to check that  
\begin{eqnarray}
\sum_{j=1}^{3}\left(
\frac{\partial 
{1\!\!\!\>\!\>{\rm l}}_{n}(|x|)}{\partial x_{j}}\right)^{2} 
\le C_{n}^{2}.
\label{eq:exp-decay-3}
\end{eqnarray}
By Proposition \ref{proposition:binding-energy} and 
(\ref{eq:exp-decay-1}) -- (\ref{eq:exp-decay-3}),  
we have 
\begin{eqnarray}
&{}&
\| G_{N_{0},\varepsilon}\otimes I\psi_{\kappa}
\|_{\mathcal H}^{2} 
\label{eq:exp-decay-4} \\ 
\nonumber 
&\le&   
C_{N_{0}}^{2}
e^{6C_{0}N_{0}}\left\{ 
|E_{\mathrm {at}}| - \sup_{N_{0}<|x|}|V(x)| 
- C_{0}^{2}\right\}^{-1}.   
\end{eqnarray}

Let $dE_{|x|}(\xi)$ be the spectral measure 
of the multiplication operator $|x|$, i.e., 
the spectral representation of $|x|$ by 
$dE_{|x|}(\xi)$ is 
\begin{eqnarray*}
|x| = \int_{0}^{\infty}\xi dE_{|x|}(\xi).
\end{eqnarray*}
Then, by Lebesgue's monotone convergence 
theorem and (\ref{eq:exp-decay-4}), 
we have 
\begin{eqnarray} 
&{}& 
C_{N_{0}}^{2}e^{6C_{0}N_{0}}
\left\{
|E_{\mathrm {at}}| - \sup_{N_{0}<|x|}|V(x)| 
- C_{0}^{2}
\right\}^{-1} 
\label{eq:exp-decay-5} \\ 
\nonumber 
&\ge& 
\lim_{\varepsilon\downarrow 0}
\| G_{N_{0},\varepsilon}\otimes I\psi_{\kappa}
\|_{\mathcal H}^{2} \\ 
\nonumber 
&=& 
\int_{0}^{\infty}
(1 - {1\!\!\!\>\!\>{\rm l}}_{N_{0}}(\xi))^{2}
e^{2C_{0}\xi}d\| 
E_{|x|}(\xi)\otimes I\psi_{\kappa}\|_{\mathcal H}^{2} \\ 
\nonumber 
&=& 
\| \left( 1 - 
{1\!\!\!\>\!\>{\rm l}}_{N_{0}}(|x|)\right) 
e^{C_{0}|x|}\otimes I\psi_{\kappa}
\|_{\mathcal H}^{2}
\end{eqnarray}
with $\psi_{\kappa} \in  
D\left(\left( 1 - 
{1\!\!\!\>\!\>{\rm l}}_{N_{0}}(|x|)\right) 
e^{C_{0}|x|}\otimes I\right)$. 
Moreover, since 
$|{1\!\!\!\>\!\>{\rm l}}_{N_{0}}(|x|)
e^{C_{0}|x|}| \le  e^{3C_{0}N_{0}}$,   
we have  
\begin{eqnarray}
\| {1\!\!\!\>\!\>{\rm l}}_{n}(|x|) 
e^{C_{0}|x|}\otimes I\psi_{\kappa}
\|_{\mathcal H}
\le 
e^{3C_{0}N_{0}}
\label{eq:exp-decay-6}
\end{eqnarray} 
with $\psi_{\kappa} \in  
D\left({1\!\!\!\>\!\>{\rm l}}_{N_{0}}(|x|)
e^{C_{0}|x|}\otimes I\right)$. 
Therefore, our statement that 
$\psi_{\kappa} 
\in  D(e^{C_{0}|x|})$ and 
(\ref{eq:new-1}) follows from (\ref{eq:exp-decay-5}) 
and (\ref{eq:exp-decay-6}).
\qed

This exponential decay immediately implies 
the following. 

\begin{proposition}[finite uncertainty of 
position in ground state]
\label{prp:|x|^{2}-domain-2}
Let $V$ be in class (C2) 
and $\kappa$ satisfy $0 \le \kappa < \Lambda$. 
If $H_{\kappa}^{\mbox{\rm {\tiny N}}}$ has 
a ground state $\psi_{\kappa}$, then 
$\psi_{\kappa} \in  D(x^{2}\otimes I)$.
Moreover, $\sup_{0<\kappa<\Lambda}
\| |x|\otimes I \psi_{\kappa}\|_{\mathcal H} 
< \infty$, provided that $\psi_{\kappa}$ 
exists for $0 < \kappa < \Lambda$. 
\end{proposition}

\proof 
We have only to note the following. 
There exists $R_{0} > 0$ such that 
$r \le e^{C_{0}r} + R_{0}$ 
for every $r \ge 0$.
\qed

\qquad 

\underbar{\it Proof of Theorem \ref{crl:2}}: 
Theorem \ref{crl:2} follows from Propositions 
\ref{prp:|x|^{2}-domain-1} and \ref{prp:|x|^{2}-domain-2} 
and Theorem \ref{theorem:2}. 

\qquad 
 
\underbar{\it Proof of Theorem \ref{crl:1}}: 
We note first that there exists a ground state $\psi_{\kappa}$ 
for $|\textsl{q}| < \textsl{q}_{\Lambda}$ 
and $0 < \kappa < \Lambda$ by 
Proposition \ref{proposition:gs-existence}. 
Then, Theorem \ref{crl:1} follows from Propositions 
\ref{prp:|x|^{2}-domain-1} and \ref{prp:|x|^{2}-domain-2} 
and Theorem \ref{theorem:1}.

\hfill\break 
{\large {\bf Acknowledgement}} 
\hfill\break  
The author thanks H. Spohn for hospitality 
at Technische Universit\"{a}t M\"{u}nchen. 
He also thanks V. Betz, F. Hiroshima, and J. L\H{o}rinczi 
for giving him useful comments about their results 
in Munich. 
He is grateful to V. Bach for hospitality 
at Johannes Gutenberg Universit\"{a}t Mainz. 
It is his pleasure to thank to M. Griesemer 
for giving advice on spatial localization, 
and to Z. Ammari and A. Pizzo for discussing 
Nelson's model in Mainz.   
He is grateful to A. Arai for his comments 
on this paper, and to J. Derezi\'{n}ski and C. G\'{e}rard 
for information about their recent results.  
He is deeply grateful to the referee for 
thoughtful comments, which helped him to correct many errors 
in the original manuscript and to reformulate 
Theorems \ref{theorem:2} and \ref{crl:2} 
in a general framework. 
In particular, the author could notice the physical image 
lying idle in the original manuscript, which is stated 
in \S \ref{section:Intro}, through the communications 
with the referee. 
Based on this image he could complete the reformations 
of them.


\begin{thebibliography}{10}


\bibitem{ammari}
Z.~Ammari, 
Asymptotic completeness for a renormalized nonrelativistic 
Hamiltonian in quantum field theory: The Nelson model,  
\textit{Math. Phys., Anal. Geom.} 
\textbf{3} (2000), 217--285. 


\bibitem{arai} 
A.~Arai, 
Ground state of the massless Nelson model without infrared cutoff 
in a non-Fock representation,   
\textit{Rev. Math. Phys.} \textbf{13} (2001), 1075--1094.   

\bibitem{arai-text} 
A.~Arai, 
Fock Spaces and Quantized Field. II (in Japanese).   
Nihon-hy\={o}ron-sha, Tokyo, 2000.   

\bibitem{ah1}
A.~Arai and M.~Hirokawa,    
On the existence and uniqueness of ground state 
of a generalized spin-boson model,  
\textit{J. Funct. Anal.} 
\textbf{151} (1997), 455--503.   

\bibitem{ah2}
A.~Arai and M.~Hirokawa,    
Ground states of a general class of quantum field Hamiltonians,  
\textit{Rev. Math. Phys.} 
\textbf{12} (2000), 1085--1135.   


\bibitem{ahh}
A.~Arai, M.~Hirokawa, and F.~Hiroshima,   
On the absence of eigenvectors of Hamiltonians in 
a class of massless quantum field models without 
infrared cutoffs,  
\textit{J. Funct. Anal.} 
\textbf{168} (1999), 470--497.   

\bibitem{ahh2}
A.~Arai, M.~Hirokawa, and F.~Hiroshima,   
Regularity of ground states in quantum field models, 
preprint 2004, arXiv:math-ph/0409055.   


\bibitem{bfs2}
V.~Bach, J.~Fr\"ohlich, and I.~M.~Sigal,   
Spectral analysis for systems of atoms and molecules 
coupled to the quantized radiation field,  
\textit{Commun. Math. Phys.} 
\textbf{207} (1999), 249--290.  

\bibitem{betz}
V.~Betz, F.~Hiroshima, J.~L\H orinczi, R.~A.~Minlos 
and  H.~Spohn,   
Gibbs measure associated with particle-field system,    
\textit{Rev. Math. Phys.} 
\textbf{14} (2002), 173--198.

\bibitem{bn}
F.~Bloch and A.~Nordsieck,   
Notes on the radiation field of the electron,  
\textit{Phys. Rev.} \textbf{52} (1937), 54--59.  


\bibitem{dg} 
J.~Derezi\'{n}ski and C.~G\'{e}rard, 
Scattering theory of infrared divergent Pauli-Fierz 
Hamiltonians, 
\textit{Ann. H. Poincar\'e} \textbf{5} (2004), 523--577.  


\bibitem{fr} 
J.~Fr\"{o}hlich, 
On the infrared problem in a model of scalar electrons and 
massless, scalar bosons,   
\textit{Ann. Inst. H. Poincar\'{e}} \textbf{19} (1973), 1--103.   


\bibitem{gerard}
C.~G\'erard,   
On the existence of ground states 
for massless Pauli-Fierz Hamiltonians,    
\textit{Ann. H. Poincar\'e} \textbf{1} (2000), 443--459.  

 
\bibitem{gll} 
M.~Griesemer, E.~H.~Lieb, and M.~Loss, 
Ground states in non-relativistic quantum electrodynamics, 
\textit{Invent. Math.} \textbf{145} (2001), 557--595. 

\bibitem{griesemer} 
M.~Griesemer,  
Exponential decay and ionization thresholds 
in non-relativistic quantum electrodynamics,  
\textit{J. Funct. Anal.} \textbf{210} (2004), 321--340. 

\bibitem{hhs2}
C.~Hainzl, M.~Hirokawa, and H.~Spohn,   
Binding energy for hydrogen-like atom 
in the Nelson model without cutoffs,  
\textit{J. Funct. Anal.} {\bf 220} (2005), 424--459.

\bibitem{hirokawa-nelson'}
M.~Hirokawa,    
Mathematical Addendum for ``Infrared Catastrophe 
for Nelson's Model'' (mp{\_}arc 03-512),  
preprint 2003, mp{\_}arc 03-551. 

\bibitem{hirokawa-e}
M.~Hirokawa, 
Recent developments in mathematical methods 
for models in nonrelativistic quantum 
electrodinamics,  
\textit{in A Garden of Quanta. 
Essays in Honor of Hiroshi Ezawa}, 
eds., J. Arafune, A. Arai, M. Kobayashi, 
K. Nakamura, T. Nakamura, I. Ojima, 
N. Sakai, A. Tonomura, and K. Watanabe, 
(2003), 209--242, World Scientific.


\bibitem{hirokawa-IR}
M.~Hirokawa,    
A Mathematical Mechanism of Infrared Catastrophe,  
preprint 2004, mp{\_}arc 04-83, arXiv:math-ph/0403008. 


\bibitem{hhs}
M.~Hirokawa, F.~Hiroshima, and H.~Spohn,   
Ground state for point particles interacting through 
a massless scalar bose field,  
\textit{Adv. Math.} {\bf 191} (2005), 339--392.

\bibitem{hiroshima}
F.~Hiroshima,    
Multiplicity of Ground States in Quantum Field Models: 
Applications of Asymptotic Fields,  
\textit{J. Funct. Anal.} {\bf 224} (2005), 431--470.


\bibitem{hk} 
R.~H{\o}egh-Krohn,   
Asymptotic fields in some models 
of quantum field theory, I,    
\textit{J. Math. Phys.} \textbf{9} (1967), 2075--2080.  

\bibitem{km}
Y.~Kato and N.~Mugibayashi, 
Regular perturbation and asymptotic limits 
of operators in quantum field theory,
{\it Prog. Theor. Phys.} {\bf 30} (1963), 103--133. 




\bibitem{lorinczi}
J.~ L\H{o}rinczi, R.~A.~Minlos and H.~Spohn,  
The infrared behaviour in Nelson's model 
of a quantum particle coupled to a massless scalar field,   
\textit{Ann. Henri Poincar\'{e}} \textbf{3} (2002), 269--295.  



\bibitem{nelson}
E.~Nelson, 
Interaction of nonrelativistic particles 
with a quantized scalar field,   
\textit{J. Math. Phys.} \textbf{5} (1964), 1190--1197.   
 



\bibitem{pf}
W. Pauli and M. Fierz,   
Zur Theorie der Emission langwelliger Lichtquanten,    
\textit{Nuovo Cimento} \textbf{15} (1938), 167--187.   

\bibitem{pi}
A. Pizzo,   
One particle (improper) states and scattering states 
in Nelson's massless model,  
\textit{Ann. Henri Poincar\'{e}} \textbf{4} (2003), 439--486.    

 
\bibitem{rs1}
M.~Reed and B.~Simon, 
\textit{Methods of Modern Mathematical Physics I: Functional Analysis},  
Academic Press, San Diego, 1980.  

 
\bibitem{rs2}
M.~Reed and B.~Simon, 
\textit{Methods of Modern Mathematical Physics II: 
Fourier Analysis, Self-adjointness},  
Academic Press, San Diego, 1980.  

\bibitem{rs4}
M.~Reed and B.~Simon, 
\textit{Methods of Modern Mathematical Physics IV: 
Analysis of Operators},  
Academic Press, San Diego, 1978.  


\bibitem{spohn}
H.~Spohn,   
Ground state of quantum particle coupled to 
a scalar boson field,    
\textit{Lett. Math. Phys.} \textbf{44} (1998), 9--16.   


\end{thebibliography}
\end{document}